\documentclass[journal,12pt,onecolumn]{IEEEtran}

\usepackage[pdftex]{graphicx}

\usepackage{fullpage}
\usepackage{cite}
\usepackage[cmex10]{amsmath}
\usepackage{algorithmic}
\usepackage{mdwmath}
\usepackage{mdwtab}
\usepackage{subfigure}

\newtheorem{theorem}{Theorem}[section]
\newtheorem{lemma}[theorem]{Lemma}

\newenvironment{proof}[1][Proof]{\begin{trivlist}
\item[\hskip \labelsep {\bfseries #1}]}{\end{trivlist}}

\newcommand{\qed}{\nobreak \ifvmode \relax \else
      \ifdim\lastskip<1.5em \hskip-\lastskip
      \hskip1.5em plus0em minus0.5em \fi \nobreak
      \vrule height0.75em width0.5em depth0.25em\fi}

\begin{document}

%
\title{A New Distributed Topology Control\\ Algorithm for Wireless
  Environments\\ with Non-Uniform Path Loss\\ and Multipath Propagation}

\author{\IEEEauthorblockN{Harish Sethu and Thomas Gerety}\\
\IEEEauthorblockA{Department of Electrical and Computer Engineering\\
Drexel University\\
Philadelphia, PA 19104-2875\\
Email: \{sethu, thomas.gerety\}@drexel.edu}
}

\maketitle
\thispagestyle{empty}
~\vskip 0.5in
\begin{abstract}
Each node in a wireless multi-hop network can adjust the power level
at which it transmits and thus change the topology of the network
to save energy by choosing the neighbors with which it directly
communicates. Many previous algorithms for distributed topology
control have assumed an ability at each node to deduce some
location-based information such as the direction and the distance of
its neighbor nodes with respect to itself. Such a deduction of
location-based information, however, cannot be relied upon in real
environments where the path loss exponents vary greatly leading to
significant errors in distance 
estimates. Also, multipath effects may result in different signal
paths with different loss characteristics, and none of these paths may
be line-of-sight, making it difficult to estimate the direction of a
neighboring node. In this paper, we present Step Topology Control
(STC), a simple distributed topology control algorithm which reduces
energy consumption while preserving the connectivity of a heterogeneous
sensor network without use of any location-based information. The STC
algorithm avoids the use of GPS devices and also makes no assumptions
about the distance and direction between neighboring nodes. We show
that the STC algorithm achieves the same or better order of
communication and computational complexity when compared to other
known algorithms that also preserve connectivity without the use of
location-based information. We also present a detailed
simulation-based comparative analysis of the energy savings and 
interference reduction achieved by the algorithms. The results show
that, in spite of not incurring a higher communication or
computational complexity, the STC algorithm performs better than other
algorithms in uniform wireless environments and especially better when
path loss characteristics are non-uniform.
\end{abstract}

\newpage
\section{Introduction}
In a multi-hop wireless sensor network, a node communicates with another node
across one or more consecutive wireless links with messages possibly
passing through intermediate nodes. The topology of such a network
can be viewed as a graph with an edge connecting any pair of nodes
that can communicate with each other directly without going 
through any intermediate nodes. Each node in such a network can choose
its own neighbors and thus control the topology by changing the power at
which it makes its transmissions or, in the case of nodes capable of
directional transmissions, by also changing the set of directions in
which it will allow transmissions. The goal of such topology control
is to employ algorithms that each node can execute in a distributed
manner for the purposes of reducing energy consumption, maintaining
connectivity, and increasing network lifetime and/or capacity.

In recent years, a large number of topology control algorithms have
been proposed and studied for a diverse set of goals
\cite{San2006}. Early work on topology control assumed that accurate
location information about its neighbors will be available to the
nodes, such as through the use of GPS devices
\cite{LiHou2005-1195,RodMen1999,LiHal2001,JiaRaj2003,LiSon2005}. This assumption
adds to the expense of the nodes and also results in high delays due
to the acquiring and tracking of satellite signals. Also, one cannot rely on
GPS in many real application environments such as inside buildings or
thick forests. Some other topology control protocols that preserve
connectivity rely on the more likely ability of a node to estimate the
distance and direction to its neighbors. For example, in the cone-based
distributed topology control (CBTC) algorithms, a node $u$ transmits
with the minimum power $p_{u,\alpha}$ required to ensure that there is
some node it can reach within every cone of degree $\alpha$ around
$u$ \cite{LiHal2005}. Assuming a specific loss
propagation model, the Euclidean distance to a neighbor can be deduced
with knowledge of the power at which a transmission is made by a
neighbor and the power at which the signal is received. The direction
of a neighbor with respect to itself can be deduced from the angle of
arrival of a signal. 

Wireless communication, however, is often characterized by the
phenomenon of multipath propagation wherein a signal reaches the
receiving antenna via two or more paths \cite{PucHae2006}. In addition, there are several
other kinds of radio irregularities that have an impact on the
topology control algorithms \cite{ZhoHe2006}. The different paths, with
differences in delay, attenuation, and phase shift, make it difficult
for the receiving node to deduce its distance from the sender and the
direction of the sender. In this paper, we focus on the design 
of topology control algorithms that work without the use of any
location-based information so that they can be employed in the
presence of multipath propagation or when path loss exponents are
non-uniform in the region of interest. More
specifically, we focus on connectivity-preserving algorithms that make
{\em no} assumptions about the availability of GPS devices in nodes
and {\em no} assumptions on the ability of a receiving node to deduce
either the distance or the direction of the sender. Besides
accommodating the environmental causes of radio irregularities, we
seek to also meet the requirements of a heterogeneous sensor network
where (a) different nodes may have different maximum transmission
powers, and (b) variations in node/antenna configurations may lead to
different reception thresholds in different directions.

\subsection{Problem Statement}
\label{problem_statement}

Assume that each node $u \in V$ is
associated with a certain maximum power $P_u$ with which it is capable
of making an omni-directional transmission (for ease of discussion, we
use omni-directional transmissions but our problem statement and the
proposed algorithm can be readily adapted for directional
transmissions). Note that we allow $P_u$ to be different for different
nodes, allowing a heterogeneous sensor network environment as in
\cite{LiHou2005-1313}. Consider the nodes in the network as vertices of a 
directed graph $G_{\mathrm max}= (V,E_{\mathrm max})$ 
in which nodes $u$ and $v$ are connected by a directed edge from $u$ to
$v$ if and only if (a) an omni-directional transmission from $u$ at
its maximum power $P_u$ can directly reach $v$, and (b) an
omni-directional transmission from $v$ at 
its maximum power $P_v$ can directly reach $u$. Note that if $(u,v)
\in E_{\mathrm max}$, then $(v,u) \in E_{\mathrm max}$. Many widely
deployed MAC and address resolution protocols in wireless networks not only assume bidirectional
links but also assume two-way handshakes and acknowledgments
\cite{IEEE802.11std}. Therefore, with bidirectional communication
assumed between directly communicating nodes,
$G_{\mathrm max}$ represents a realistic communication topology at maximum node
powers. 

Given $G_{\mathrm max} = (V,E_{\mathrm max})$, the goal of topology
control in this paper can be thought of as a multi-objective
optimization problem where we seek a new weighted directed graph $G =
(V,E)$, $E \subseteq E_{\mathrm max}$. We define $P_G(u)$ as the minimum
power with which node $u$ should make an omni-directional transmission
so as to reach all of its neighbors in graph $G$. Let $C_G(u)$ denote
the energy cost of a transmission by $u$ at power
$P_G(u)$.\footnote{Note that the power at which a node makes its
  transmissions is the energy consumed in the transmission per unit
  time. When the context is a single node, reducing power and reducing
  energy become the same goal because when we reduce transmission
  power at a given node, we also reduce energy consumption at the
  node. The energy consumption by the overall network, however, is not
  always reduced when the transmission powers at the nodes are
  reduced. In this paper, we use the goals of reducing the energy cost
  and reducing the transmission power interchangeably when the context
  is limited to the transmissions by a single node. In the context of
  an entire network, we speak of the goal of reducing energy costs.}
Assign weight $C_G(u)$ to each directed edge $(u,v) \in E$ (all edges
starting from a node have the same weight since nodes in a sensor
network broadcast messages at a certain constant power level
determined by the topology control algorithm). Let $C_G(s \rightarrow 
t, {\mathrm Energy})$ denote the sum of $C_G(i)$ for each transmitting
node $i$ in the minimum-energy path from node $s$ to node $t$ in the
graph $G$ (this represents the minimum energy cost for a message to
traverse from a source $s$ to a destination $t$).

The multi-objective problem is to find $G$ with the following two objectives:
\begin{enumerate}
\item Minimize the average power, computed over all nodes, required
  in  an omni-directional transmission by a node $u$ to reach all of
  its neighbors in $G$. This objective can be expressed as:
\[
\min \left[ \sum_{u \in V} C_G(u) \right]
\]
\item Minimize the average energy cost, computed over all
  source-destination pairs, in a minimum-energy path from a
  source node to a destination node. This objective can be expressed as: 
\[
\min \left[ \sum_{s, t \in V} C_G(s \rightarrow t, {\mathrm Energy}) \right]
\]
\end{enumerate}
under the following constraints:
\begin{enumerate}
\item $E \subseteq E_{\mathrm max}$.
\item If $(u,v) \in E$, then $(v, u) \in E$, as is generally expected by MAC layer protocols.
\item If there exists a path between $u$ and $v$ in $G_{\mathrm max}$, then
  there also exists a path between $u$ and $v$ in $G$ (preserves connectivity).
\end{enumerate}

The two objectives of minimizing the average transmission power
of a node and minimizing the average energy cost along a minimum-energy
path often work against each other. For example, minimizing average
transmission power can lead to a very sparse graph with very long
energy-expensive multi-hop paths between some pairs of nodes.

\subsection{Related Work}

\begin{figure*}[!t]
\begin{center}
\framebox[15cm]{
\parbox{7cm}{
{\small
\begin{tabbing}
ww \= ww \= ww \= ww \= ww \= ww \= w \kill
\mbox{~}\\ 
\mbox{~}01: \> {\em Function} {\bf STC} at node $(u)$:\\
\mbox{~}02: \> \> $G \leftarrow (V, \phi )$ /* directed graph with
no edges */\\
\mbox{~}03: \> \> Compile {\em outTupleList}$(u)$ and {\em inTupleList}$(u)$\\
\mbox{~}04: \> \> Broadcast {\em outTupleList}$(u)$ and {\em inTupleList}$(u)$ at maximum power $P_u$\\
\mbox{~}05: \> \> Receive {\em outTupleList}$(n)$ and {\em inTupleList}$(n)$ from each neighbor $n$ in $G_{\mathrm max}$\\
\mbox{~}06: \> \> Compute {\em fPairOfPaths}$(n)$ for each $n$ two
or fewer hops away from $u$\\
\mbox{~}07: \> \> Compute {\em bPairOfPaths}$(n)$ for each $n$ two
or fewer hops away from $u$\\
\mbox{~}08: \> \> Sort {\em outTupleList}$(u)$\\
\mbox{~}09: \> \> $k \leftarrow$ degree of $u$ in $G_{\mathrm max}$\\
\mbox{~}10: \> \> do $k-1$ times\\
\mbox{~}11: \> \> \> $t( u, v ) \leftarrow$ the largest tuple in {\em outTupleList}$(u)$\\
\mbox{~}12: \> \> \> remove $t( u, v )$ from {\em outTupleList}$(u)$\\
\mbox{~}13: \> \> \> {\em vSet} $= \{ i\mbox{~}|\mbox{~}t( i, v ) \in
{\mathit inTupleList} ( v ), t(i, v) < t(u, v)\}$\\
\mbox{~}14: \> \> \> {\em NoForwardPath} $=$ True\\
\mbox{~}15: \> \> \> {\em NoBackwardPath} $=$ True\\
\mbox{~}16: \> \> \> for $n \in $ {\em vSet}\\
\mbox{~}17: \> \> \> \> $p = $ the first path in {\em fPairOfPaths}$(n)$
without $v$\\
\mbox{~}18: \> \> \> \> if $\max \{ r | r \in p \} < t(u, v)$\\
\mbox{~}19: \> \> \> \> \> {\em NoForwardPath} $=$ False\\
\mbox{~}20: \> \> \> \> \> break (out of for loop)\\
\mbox{~}21: \> \> \> \> end if\\
\mbox{~}22: \> \> \> end for\\
\mbox{~}23: \> \> \> {\em vSet} $= \{ i\mbox{~}|\mbox{~}t( v, i ) \in
{\mathit outTupleList} ( v ), t(v,i) < t(v,u)\}$\\
\mbox{~}24: \> \> \> for $n \in $ {\em vSet}\\
\mbox{~}25: \> \> \> \> $p = $ the first path in {\em bPairOfPaths}$(n)$
without $v$\\
\mbox{~}26: \> \> \> \> if $\max \{ r | r \in p \} < t(v, u)$\\
\mbox{~}27: \> \> \> \> \> {\em NoBackwardPath} $=$ False\\
\mbox{~}28: \> \> \> \> \> break (out of for loop)\\
\mbox{~}29: \> \> \> \> end if\\
\mbox{~}30: \> \> \> end for\\
\mbox{~}31: \> \> \> if {\em NoForwardPath} or {\em NoBackwardPath}\\
\mbox{~}32: \> \> \> \> add a directed edge $( u, v )$ to $G$\\
\mbox{~}33: \> \> \> end if\\
\mbox{~}34: \> \> end do\\
\end{tabbing}
} } }
\end{center}
\caption{Pseudo-code description of the Step Topology Control 
  algorithm at node $u$.} \label{fig:stc-pseudo-code}
\end{figure*}

Let $C(u,v)$ denote the minimum energy cost of a successful
transmission from $u$ to $v$. A topology control algorithm that minimizes energy consumption
will remove an edge $(u,v)$ if and only if there exists a
path between $u$ and $v$ through an intermediate set of nodes $n_1,
n_2 \dots n_k$ such that $C(u,n_1) + C(n_1, n_2) + \dots
C(n_k, v) < C(u,v)$. Accomplishing this requires a significant exchange of
information between nodes and in such a case, the topology control
algorithm is indistinguishable from a routing algorithm. As a result,
a number of distributed topology control algorithms have been proposed
where nodes rely on lesser exchange of information between neighbors
\cite{San2006}. In this subsection, we focus on the subset of these protocols that
can be adapted to work without the exchange of any location-based
information between neighboring nodes.

The KN{\small EIGH} protocol is based on determining the number of
neighbors that each node should have in order 
to achieve full connectivity with a high probability
\cite{BloLeo2003}. This protocol, however, does not guarantee 
connectivity even though it does achieve connectivity with a high
probability. The Small Minimum-Energy Communication
Network (SMECN) protocol \cite{LiHal2001} seeks to achieve a lower
energy cost while guaranteeing connectivity by removing an edge
$(u,v)$ if and only if there exists a path $u \rightarrow n
\rightarrow v$ such that $C(u,n) + C(n, v) < C(u,v)$. As proposed in
\cite{LiHal2001}, the protocol requires the use of GPS devices but the
same results can be accomplished if each node exchanges information
with each of its neighbors regarding the energy costs of reaching all of its
neighbors. When used with some of the widely available routing algorithm
implementations such as AODV or DSR that base their decisions on the
number of hops in a path rather than the total energy cost of the
path, the SMECN protocol does not necessarily result in a significant
reduction in energy consumption. As will be shown in Section
\ref{sec:Simulation}, the energy savings achieved by SMECN is not
significantly greater even if the routing algorithm were to choose a
minimum-energy path.    

The Directed Relative Neighborhood Graph (DRNG) protocol \cite{LiHou2005-1313} removes an edge
$(u,v)$ if and only if there exists a path $u \rightarrow n
\rightarrow v$ such that $\max\{ C(u,n), C(n, v) \} < C(u,v)$. In the
Directed Local Spanning Subgraph (DLSS) protocol \cite{LiHou2005-1313},
each node creates a local spanning tree from the subgraph induced by
itself and its neighbors. A node $u$ retains the edge $(u,v)$ in the
topology-controlled graph if and only if the edge $(u,v)$ exists in
the local spanning tree generated at node $u$. As proved in
\cite{LiHou2005-1313}, any edge that is removed by DRNG is also
removed by DLSS and therefore, DLSS achieves a sparser graph than
DRNG. The fewer edges in DLSS often leads to a 
lower energy consumption, but it can sometimes lead to longer paths
and therefore higher energy consumption than DRNG. A generalization of
the DRNG protocol is the XTC protocol which was conceived
independently and uses the notion of `link quality' in making its
topology control decision instead of energy costs
\cite{WatZol2004}. When link quality between two nodes is measured by
the energy cost of a transmission between them, the XTC protocol 
reduces to DRNG.

Among other attempts to accommodate the irregularities of a real
wireless environment in topology control are some that allow for
uncertainties  
in whether or not a nearby node is reachable if the distance to that
node is above a certain threshold
\cite{DamPan2006,KuhZol2003}. However,
these algorithms assume that each node can know the distances to other
nearby nodes, something that cannot be relied upon in a real
environments with multipath effects. Some other works
have also considered realistic wireless models but they have only presented
centralized algorithms \cite{BloLeo2005}.

\subsection{Contributions}

In this paper, we present the Step Topology Control (STC) algorithm
in which a node $u$ removes an edge $(u,v)$ if and only if there exists:
\begin{itemize}
\item a path with three or fewer hops from $u$ to $v$
  such that the energy cost across each hop is less than $C(u,v)$.
\item a path with three or fewer hops from $v$ to $u$
  such that the energy cost across each hop is less than $C(v,u)$.
\end{itemize}
The STC algorithm relies on
each node exchanging information with each of its neighbors regarding
the energy costs of communication to all of its neighbors. 

The STC algorithm may be seen as an extension of the DRNG algorithm to
allow a search for three-hop paths and to ensure bidirectional
communication between directly communicating nodes. However, our
implementation of the algorithm ensures that the order of communication and the
computational complexity do not increase despite a search for
three-hop paths. This search for
three-hop paths makes only a small impact on performance when the
wireless environment is uniform across the network. However, it makes a
{\em significant} impact when a more realistic scenario is assumed
with multipath propagation and other irregularities in the path loss
characteristics in the wireless environment. In such environments,
the STC algorithm achieves a much lower energy consumption
and interference than other topology control algorithms in spite of
maintaining the same or better order of communication and
computational complexity. We also show that
this improved performance of the STC algorithm exists largely
independent of the size of the network.

Section \ref{sec:STC} presents the STC algorithm along with a
pseudo-code description. Section \ref{sec:Comparison} presents an
analysis of the communication and computational complexity of the STC
algorithm and presents a comparative analysis. Section
\ref{sec:Simulation} presents several simulation-based results that
provide a thorough comparison of the energy consumption properties of the STC algorithm and other
existing topology control strategies that can be adapted to use no
location-based information. In particular, we examine the algorithms
in environments with both constant and varied path loss exponents and
study their scalability in performance as the number of nodes
increases. Section \ref{sec:Conclusion} concludes the paper with a
summary of its findings and future research directions. A proof of the
relationship between the STC algorithm and a CBTC algorithm with all
applicable optimizations is presented in the Appendix.

\section{Step Topology Control}
\label{sec:STC}

Denote by $P_{\mathrm min}(u,v)$, the minimum power necessary for a
transmission from $u$ to reach $v$, otherwise known as {\em
  transmission power threshold}. We allow that $P_{\mathrm min}(u,v)$ is not
necessarily the same as $P_{\mathrm min}(v,u)$. The basic idea behind the Step
Topology Control (STC) algorithm is to find both forward and backward
paths with three or fewer hops between $u$ and $v$ such that each hop
requires a lower energy cost than that required for an equivalent direct
transmission between $u$ and $v$. If such multi-hop paths exist between $u$
and $v$, node $u$ drops the edge $(u,v)$ from the directed graph it
generates and node $v$ similarly drops the edge $(v,u)$. 

The STC algorithm relies on being able to uniquely order the energy costs
of transmissions across nodes. Since the power levels at which
transmissions are made may take on only certain discrete values in
some systems, we add additional identifiers to permit a unique
ordering. We assume that each node $u$  
is uniquely identified by an integer ID$_u$. For each ordered pair of
nodes $u$ and $v$, we associate an ordered tuple $t(u,v) =
(t_1,t_2,t_3)$, where $t_1 = P_{\mathrm min}(u,v)$, $t_2 = \mathrm{ID}_u$, and
$t_3 = \mathrm{ID}_v$. We say that $(t_1,t_2,t_3) < (t_1^\prime,
t_2^\prime, t_3^\prime)$ if and only if (1) $t_1 < t_1^\prime$, or (2)
$t_1 = t_1^\prime$ and $t_2 < t_2^\prime$, or (3) $t_1 = t_1^\prime$,
$t_2 = t_2^\prime$, and $t_3 < t_3^\prime$. For the sake of
completeness, we define $t(u,u) < t(x,y)$ for any $u$ and $x \neq y$ since a transmission to itself
should cost less energy than a transmission to another node. Note that $t(u,v)$ and
$t(v,u)$ are strictly ordered by the above lexicographic rule and not
equal even if the minimum power required for transmission between $u$
and $v$ is the same in either direction. We call $t(u,v)$ a
transmission tuple. A path $p$ in the graph consists of an ordered
sequence of transmission tuples. Denote by {\em maxTuple}$(p)$ the
largest tuple in path $p$.

Consider any two nodes $u$ and $v$ such that $(u,v) \in E_{\mathrm max}$. We
assume that $u$ can determine the minimum power necessary for its
transmission to reach $v$ as well as the minimum power required for
a transmission from $v$ to reach itself. This is accomplished by
transmitting beacon messages at increasing powers and noting the
power at which each neighbor is first discovered. Each beacon message
carries within it the power at which it is transmitted so that the
discovered neighbor may also note the minimum power necessary for it
to be reached by a neighboring node. Each node $u$ can thus compile two
lists of transmission tuples: {\em outTupleList}$(u)$ containing
$t(u,v)$ for all $(u,v) \in E_{\mathrm max}$, and {\em inTupleList}$(u)$
containing $t(v,u)$ for all $(v,u) \in E_{\mathrm max}$. This process of
exchanging power level information is feasible in practice and is part
of many proposed energy-aware MAC layer protocols as well as topology
control algorithms such as DLSS.

Figure \ref{fig:stc-pseudo-code} presents a pseudo-code description
of the STC algorithm. Once a node $u$ compiles {\em outTupleList}$(u)$ and {\em
  inTupleList}$(u)$, it begins execution of the algorithm. Each node
$u$ first broadcasts both its {\em inTupleList} and {\em outTupleList}
at its maximum power $P_u$ to reach all of its neighbors (line 04). The node
also collects the {\em inTupleLists} and {\em outTupleLists} from each of its
neighbors (line 05). 

Given all this information about the energy costs, node $u$ computes 
two forward paths (denoted {\em fPairOfPaths}$(n)$) and two backward paths
(denoted {\em bPairOfPaths}$(n)$) for each node $n$ that is
reachable by two or fewer hops (lines 06--07). We describe below in greater detail
the construction of the forward path data structures for node $n$
(the construction of the backward pair of paths is similar). Let $p_1$
denote the first of the pair of paths in {\em fPairOfPaths}$(n)$
and let $p_2$ denote the second. We choose $p_1$ and $p_2$ 
such that {\em maxTuple}$(p_1) < $ {\em maxTuple}$(p_2) < $ {\em maxTuple}$(p)$ where $p$ is any path of two or fewer hops from
$u$ to $n$ other than $p_1$ and $p_2$. In lines 16--22, this data
structure allows the node to quickly determine if there exists a path
$s$ of three or fewer hops between $u$ and another node $v$ such that
{\em maxTuple}$(s) < t(u,v)$. The reason we need a pair of paths
instead of just one path is because one of these paths may be through
$v$ and one would not choose to replace the edge $(u,v)$ with a path that
goes through $v$. 

Node $u$ then orders the 
tuples in its {\em outTupleList}$(u)$ (line 08) and
considers each of the edges $(u,v)$ in reverse lexicographical order of the
associated tuples $t(u,v)$, i.e., the neighbor that requires the largest power
to be reached is considered first (line 11). As each neighbor $v$ is 
processed, the corresponding tuple $t(u,v)$ is removed from {\em
  outTupleList}$(u)$ (line 12). 

To determine if an edge $(u,v)$ should be removed, a node $u$ looks
for a forward path $u \rightarrow n_1 
\rightarrow n_2 \rightarrow v$, where the nodes $n_1$ and $n_2$ may or
may not be distinct (lines 13--22). The condition that the path should
satisfy is $\max\{ t(u, n_1), t(n_1, n_2), t(n_2, v)\} < t(u,v)$. The node similarly seeks to find a reverse path
via nodes $n_3$ and $n_4$ (lines 23--33) such that $\max\{ t(v, n_3), t(n_3, n_4), t(n_4, u)\} < t(v,u)$.
Note that the number of distinct nodes among $n_1$, $n_2$, $n_3$
and $n_4$ may range anywhere between 1 and 4.
If both forward and backward two- or three-hop paths are found satisfying
the desired conditions on the tuples, the edge $(u,v)$ is
removed from the graph (lines 31--33).

To determine if there exists a forward path from $u$ to $v$ satisfying the
desired conditions, node $u$ first constructs the set
{\em vSet} consisting of nodes $i$ that are neighbors of $v$
such that $t(i, v) < t(u, v)$ (line 13). Now, if there exists a path, $p$, of two or
fewer hops from $u$ to any node in {\em vSet} such that the path
is not through $v$ and {\em maxTuple}$(p) < t(u,v)$, then the
condition for the forward path is satisfied. This is determined in
lines 16--22 and the existence of a backward path is similarly
determined in lines 24--30.

\section{Comparative Analysis}
\label{sec:Comparison}

In this section, we discuss the communication and computational
complexity of the STC algorithm in comparison to other topology
control algorithms that also preserve connectivity while being
capable of operation in real environments with multipath propagation
or non-uniform path loss exponents. We also discuss any
provable relationships between STC and other algorithms regarding
the set of edges removed from $G_{\mathrm max}$ by the algorithm.

\subsection{Complexity}
\label{sec:Complexity}

We consider the complexity of the following algorithms and present a
comparison to the STC algorithm:
\begin{itemize}
\item Small Minimum Energy Communication Network (SMECN).
\item Directed Relative Neighborhood Graph (DRNG).
\item Directed Local Spanning Subgraph (DLSS).
\end{itemize}
Each of the above algorithms relies on each node first determining the
energy cost to each of its neighbors. As explained in Section
\ref{sec:STC}, this can be accomplished by transmitting beacon
messages at increasing powers and noting the power at which each
neighbor is first discovered. When beacon messages carry the power
at which they are being transmitted, each node can also learn the
energy cost of the communication from each of its neighbors to
itself. Depending on the granularity of power levels at which
transmissions can be made, a variety of strategies based on linear or
binary search methods may be employed in these steps to minimize
interference and to maximize speed of convergence to the correct power
values. We do not include this step in our complexity analysis because
it is common to all topology control protocols above and cannot be
considered a distinguishing feature of STC or any other specific
protocol. Topology control algorithms begin with the assumption that
each node can determine and will know the minimum power at which it
should transmit to reach another particular node. For the STC
algorithm, for example, we consider its complexity at node $u$ after
the compilation of {\em outTupleList}$(u)$ and {\em inTupleList}$(u)$
is complete.

In the original graph, $G_{\mathrm max}$, on which a topology control
algorithm is executed, the in-degree of a node is the same as its
out-degree (since bidirectional communication is assumed) and
therefore, the maximum out-degree ($\Delta^+$) and the maximum
in-degree ($\Delta^-$) are identical. We define the node-degree of a
node in $G_{\mathrm max}$ as the number of its neighbors that it
communicates with, i.e., its out-degree or its in-degree. In the
following, we denote the maximum node-degree of the original graph by
$\Delta = \Delta^+ = \Delta^-$. Further, in our analysis, we do not
include the ID size in computing the communication complexity. This is
because the IDs on sensor nodes are likely to be globally unique for
each node and assigned by the manufacturer, as in the case of Ethernet
cards. We expect the ID sizes, therefore, to be independent of the
size of the sensor network deployment.


\begin{theorem}
\label{theo:STC-comm-complexity}
The communication complexity of the STC algorithm
is $O( \Delta^2 )$.
\end{theorem}

\begin{proof}
For any given node $n$, the lists {\em inTupleList}$(n)$ and {\em outTupleList}$(n)$ are each of length
$O( \Delta )$. Broadcasting the lists (line 04 in
Figure \ref{fig:stc-pseudo-code}), therefore, is $O( \Delta )$ in
communication complexity and receiving the lists from each of up to
$\Delta$ neighbors (line 05) is $O( \Delta^2 )$ in communication
complexity. The overall communication complexity, therefore, is $O(
\Delta^2 )$.
\end{proof}

\begin{theorem}
\label{theo:STC-comp-complexity}
The computational complexity of the STC algorithm
is $O( \Delta^2 )$.
\end{theorem}

\begin{proof}
For each neighbor $n^\prime$ of $u$, one can process the entries in the
{\em outTupleList}$(n^\prime)$ to create the pair of paths in {\em
  fPairOfPaths}. The determination of whether or not a path to node
$n$ should be included in {\em fPairOfPaths} and whether it
should be the first or the second of the pair of paths can be made in
$O(1)$ time since it involves no more than two tuple comparisons.
Let $h$ denote the number of one-hop or two-hop paths starting from
$u$. Since $h = O( \Delta^2 )$, creating the {\em
  fPairOfPaths} will require a total time of $O( \Delta^2 )$. Creating the {\em
  bPairOfPaths} will similarly require a total time of $O( \Delta^2
)$.

Sorting of the {\em outTupleList}$(u)$ in line 08 takes time $O(
\Delta \log \Delta)$ since the size of the list is $O(\Delta)$.

{\small
\begin{table}[!t]
\begin{center}
\begin{tabular}{|l|c|c|}\hline
{\rule[-3mm]{0mm}{8mm}} Algorithm & Communication Complexity &
Computational Complexity\\
\hline \hline
{\rule[-2mm]{0mm}{6mm}} SMECN & $O( \Delta^2 )$ & $O( \Delta^2 )$ \\ \hline
{\rule[-2mm]{0mm}{6mm}} DRNG & $O( \Delta^2 )$ & $O( \Delta^2 )$ \\ \hline
{\rule[-2mm]{0mm}{6mm}} DLSS & $O( \Delta^2 )$ & $O( \Delta^2 \log \Delta )$ \\ \hline
{\rule[-2mm]{0mm}{6mm}} STC & $O( \Delta^2 )$ & $O( \Delta^2 )$ \\ \hline 
\end{tabular}
\end{center} 
\caption{A Comparison between Topology Control Algorithms}
\label{fig:complexity_table}
\end{table}
}

We now show that the $2h$ pairs of paths created above can allow the inner {\tt for}
loops of lines 16--22 and 24--30 to complete in time $O( \Delta )$. It
is readily verified that each iteration of the {\tt for} loops
completes in $O(1)$ time (lines 17--21 and 25--29). Since the size of
the {\em vSet} is $O( \Delta )$, these inner {\tt for} loops execute in
$O( \Delta )$ time. The only other non-loop component inside the outer {\tt do}
loop between lines 10--34 that takes more than $O(1)$ time is the
creation of the {\em vSet} itself, which takes
$O(\Delta)$ time. Since the {\tt do} loop iterates $k-1$ times where $k
\leq \Delta$, the entire {\tt do} loop completes in time $O( \Delta^2
)$.

The overall computational complexity of the STC algorithm, therefore,
is $O( \Delta^2 )$.

\end{proof}

Table \ref{fig:complexity_table} presents a comparison of
the communication and computational complexities of topology control
algorithms that preserve connectivity and which do not employ the
exchange of any location-based information between neighbors. All
involve a communication complexity of $O( \Delta^2 )$ since they all
require that each node collect a list of energy costs from each of its
neighbors. In the computational complexity analysis of DLSS, we
consider the number of nodes in the local graph as $O( \Delta )$ and
the number of edges as $O( \Delta^2 )$. The computation of the local
minimum spanning tree, assuming Kruskal's algorithm \cite{Kru1956}, is
$O( \Delta^2 \log \Delta )$.

\subsection{Set of edges removed}
\label{sec:Edges}

The SMECN, DRNG and DLSS algorithms allow directed edges in the graph
they generate. Since STC assumes bidirectional links motivated by the need of
real MAC layer protocols, in order to make a fair comparison, we will
assume that energy costs are the same in both forward and reverse
directions (so that SMECN, DRNG and DLSS will also lead to
only undirected edges in the graphs they generate). Under this
assumption, the STC algorithm removes any edge from the original graph
that is also removed by SMECN or DRNG. The SMECN algorithm removes an
edge $(u,v)$ if there exists a two-hop path from $u$ to $v$ 
through $n$ such that $C(u,n) + C(n,v) < C(u,v)$. Therefore, we have
$C(u,n) < C(u,v)$ and $C(n,v) < C(u,v)$ which would be the condition
that will lead the DRNG or STC algorithm to also remove edge
$(u,v)$. Assuming that the energy costs are the same in both forward
and reverse directions, the STC algorithm becomes a simple extension
of the DRNG algorithm, and therefore, it is easily argued that it will
remove all edges that are removed by DRNG. Thus, the STC algorithm
yields a lower energy cost per transmission at any given node
than either SMECN or DRNG under these conditions. However, the STC
algorithm does not always remove an edge that would be removed by
DLSS. For a comparative analysis with the DLSS algorithm, we rely on
the simulation results presented in the next section.

It is of interest to note that the STC algorithm is related to the
OPT-CBTC($5\pi/6$) algorithm (which is the CBTC($5\pi/6$) algorithm with all
applicable optimizations). The STC algorithm removes all edges that
would be removed by OPT-CBTC($5\pi/6$). As a result of this
relationship, proved in Appendix A, the STC algorithm exhibits some of
the same angular properties as OPT-CBTC($5\pi/6$).

Since DRNG, DLSS and STC do not necessarily reduce energy costs in the
path from $u$ to $v$ each time they remove an edge $(u,v)$,
rare pathological cases are possible where these algorithms
significantly increase energy costs. Consider an edge $(u,v)$
corresponding to an energy cost of $C(u,v)$. It is possible that the
STC algorithm removes this edge because there exist three other edges,
each corresponding to an energy cost of $C(u,v)- \epsilon$, where
$\epsilon$ is an infinitesimal quantity (thus almost tripling the energy
cost between $u$ and $v$). These three edges, in turn may be removed
because each can be replaced by three edges corresponding to energy
costs of $C(u,v)- 2\epsilon$, which now multiplies the total energy
cost between $u$ and $v$ by nine. Such scenarios are rare in real
situations and our simulation results, described in the next section,
show that when nodes are scattered randomly in a region, the STC
algorithm actually performs significantly better than the other
algorithms considered in this paper.

\section{Simulation Results}
\label{sec:Simulation}

In this section, we present a simulation-based comparative study of the following
distributed topology control algorithms:
\begin{itemize}
\item Step Topology Control (STC), the algorithm presented in this paper.
\item OPT-CBTC($5\pi/6$), as a representative CBTC algorithm and
  because of its relationship to the STC algorithm.
\item Small Minimum Energy Communication Network (SMECN).
\item Directed Relative Neighborhood Graph (DRNG).
\item Directed Local Spanning Subgraph (DLSS).
\end{itemize}
The XTC protocol uses the notion of `link quality' in generating the
output topology \cite{WatZol2004}. When the link quality is measured
by the energy cost of a transmission across the link, the XTC protocol
reduces to the DRNG protocol. Therefore, the XTC protocol is not
separately included in these comparisons.

In addition to the above, we use the following two additional
algorithmic metrics as references on the performance bounds of
topology control algorithms:
\begin{itemize}
\item {\em Minimum Spanning Tree (MST):} The MST achieves the first stated
  objective of the topology control problem described in
  Section~\ref{problem_statement}. The MST preserves connectivity
  while minimizing the average power (computed over all nodes)
  with which a node makes its transmissions.
\item {\em MinReach:} In MinReach, each transmission from a node to
  one of its neighbors uses exactly the {\em minimum} energy required
  to reach that particular neighbor (as opposed to each node making
  all its transmissions to any neighbor at the same power level). The total
  energy cost of transmissions from a source to a destination along
  the minimum energy path using MinReach is a lower bound (though, not
  a tight lower bound) on the total energy cost along a path between
  the source-destination pair\footnote{Some other topology control algorithms, such as a
    distributed algorithm for constructing a Gabriel Graph
    \cite{SonWan2004}, cannot serve as optimal solutions or lower bounds
    in our case where different nodes may have different maximum
    transmission powers and where path loss exponents may vary in
    different directions for the same transmitting node.}.  
\end{itemize}

\subsection{Metrics}

Assume $P_u = P$ for all nodes $u$ (recall
that $P_u$ denotes the maximum power with which a node $u$ can
transmit). If the power $P$ is assumed to be an arbitrary quantity in
our simulations, the reduction in energy achieved by topology control
algorithms can be misleading (if $P$ is arbitrarily large, the
reduction will be similarly large). Therefore, in our simulations, we
assume the smallest possible value of $P$ so that the original graph
used as the input to the topology algorithm is connected. This offers
a standardized approach to estimating the effectiveness of topology
control algorithms so that arbitrarily large reductions in energy
consumption cannot be claimed by topology control algorithms by simply
using a dense highly-connected $G_{\mathrm max}$. For each 
network, the baseline for our comparisons is the {\em initial} graph,
$H$, defined as follows. Consider graph $G$ generated by creating an
edge $(u,v)$ from $u$ to $v$ if and only if a transmission from $u$ at
power $p$ can reach $v$ and a transmission from $v$ to $u$ at power
$p$ can reach $u$. Note that, when $p=P$, $G = G_{\mathrm max}$. Let
$P_H$ denote the minimum value of $p$ at which $G$ is connected. The
graph $G$ generated by each node transmitting at power $P_H$ is the
initial graph, $H$.  

Given a graph, $T^\prime$, generated by the execution of the topology control
algorithm, we define $P_{T^\prime}(u)$ as the minimum power with
which node $u$ should make an omni-directional transmission so as to
reach all of its neighbors in graph $T^\prime$. Let $C_{T^\prime}(u)$ denote the energy cost of a 
transmission by $u$ at power $P_{T^\prime}(u)$.  

Given the graph $T^\prime$ above, we define a new graph $T$ as follows: a node $u$ is connected by an
edge to $v$ in $T$ if and only if $C_{T^\prime}(u) > C(u,v)$ or $C_{T^\prime}(v) >
C(v,u)$. Thus, given omni-directional transmissions, $T$
represents the graph that is actually relevant for performance comparisons and, especially,
interference comparisons. $T$, as opposed to $T^\prime$, also enforces the ability for
bidirectional transmissions between communicating nodes. We call $T$ the {\em cover graph}
generated by the topology control algorithm. All of our simulation
results use the cover graph. 

Define $P_T(u)$ as the minimum power with
which node $u$ should make an omni-directional transmission so as to
reach all of its neighbors in the cover graph $T$. Let $C_T(u)$ denote the energy cost of a 
transmission by $u$ at power $P_T(u)$. Note that $P_T(u) \geq
P_{T^\prime}(u)$ and $C_T(u) \geq C_{T^\prime}(u)$. In our
simulations, we use $P_T(u)$ as the power at which node $u$ makes all
its transmissions after the execution of a topology control algorithm
generating cover graph $T$.

$P_T(u)/P_H$, denoted by $P_{T/H}(u)$, is the ratio of
the power at which node $u$ transmits after the execution of the
topology control algorithm that generates cover graph $T$ and the power at which it transmits in
the initial graph, $H$. With this normalization to the energy costs in
the initial graph, this ratio captures the energy savings per
transmission due to the topology control algorithm. 

Let $C_T(u \rightarrow v, {\mathrm Hops})$ denote the sum of $C_T(i)$ for each
transmitting node $i$ in the minimum-hop path from $u$ to $v$ in cover graph $T$. $C_T(u
\rightarrow v, {\mathrm Hops})/C_H(u \rightarrow v, {\mathrm Hops})$, denoted by $C_{T/H}(u
\rightarrow v, {\mathrm Hops})$, is the ratio of the energy cost along
the minimum-hop path from
$u$ to $v$ in cover graph $T$ generated by the topology control algorithm 
and the corresponding cost along the minimum-hop path from $u$ to $v$
in the initial graph, $H$. This ratio captures the energy savings
along a path due to the topology control algorithm. In our simulation
experiments, we examine both the minimum-energy paths and the
minimum-hop paths. The ratio for the minimum-energy paths is computed
similarly as above and is denoted by $C_{T/H}(u
\rightarrow v, {\mathrm Energy})$.

\begin{figure*}[!t]
\begin{center}
    \subfigure[{Average $P_{T/H}(u)$ over all $u$.}]{
        \label{fig:Fig2_avgOmniPower}
        \includegraphics[width=3in]{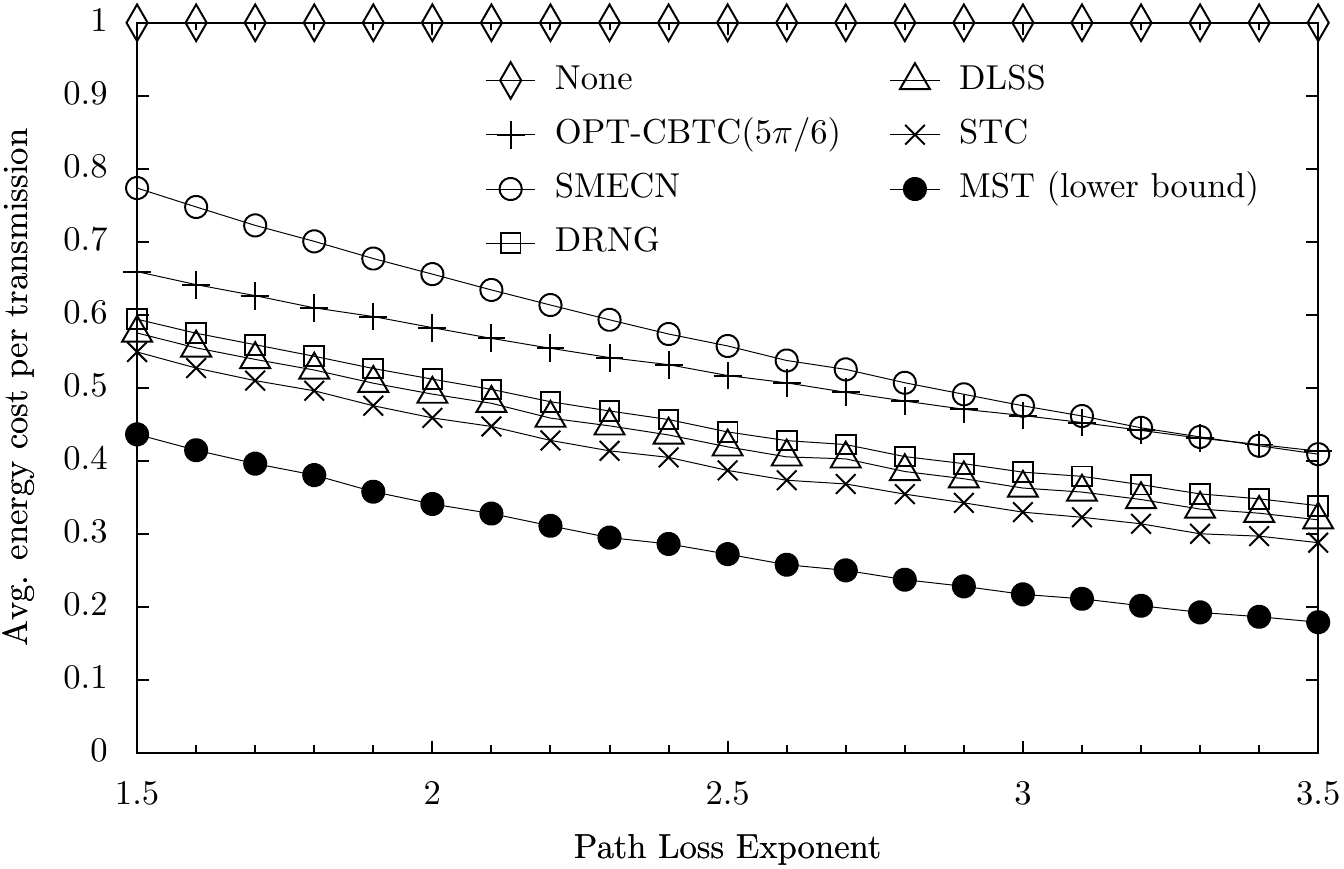}
        }
    \hskip0.1in 
    \subfigure[{$C_{T/H}(u \rightarrow v, {\mathrm Energy})$ averaged over all
      pairs of nodes $u$ and $v$.}]{
        \label{fig:Fig2_avgOptPathOmniPower}
        \includegraphics[width=3in]{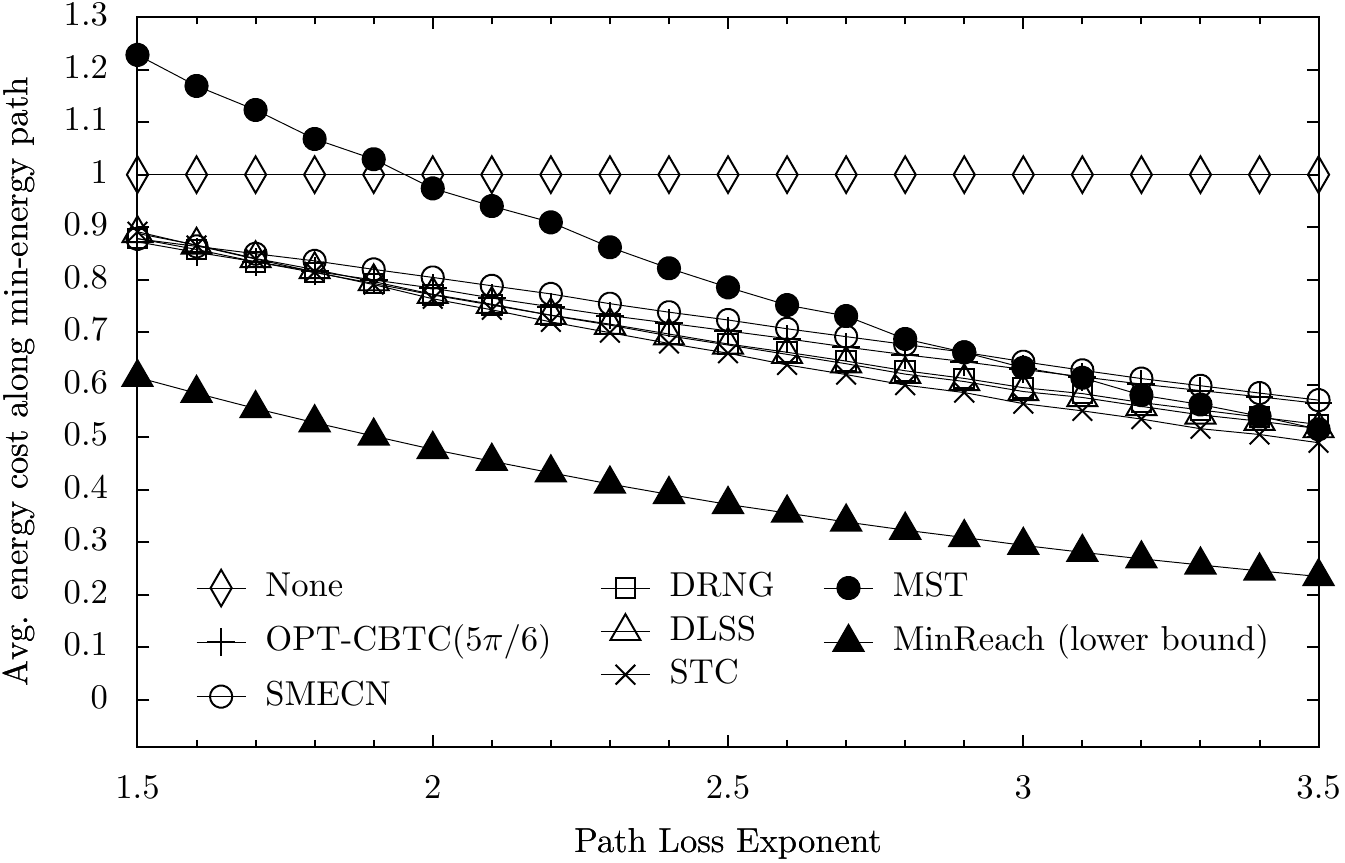}
        }\\
    \subfigure[{Average node-degree.}]{
        \label{fig:Fig2_avgOmniNodeDegree}
        \includegraphics[width=3in]{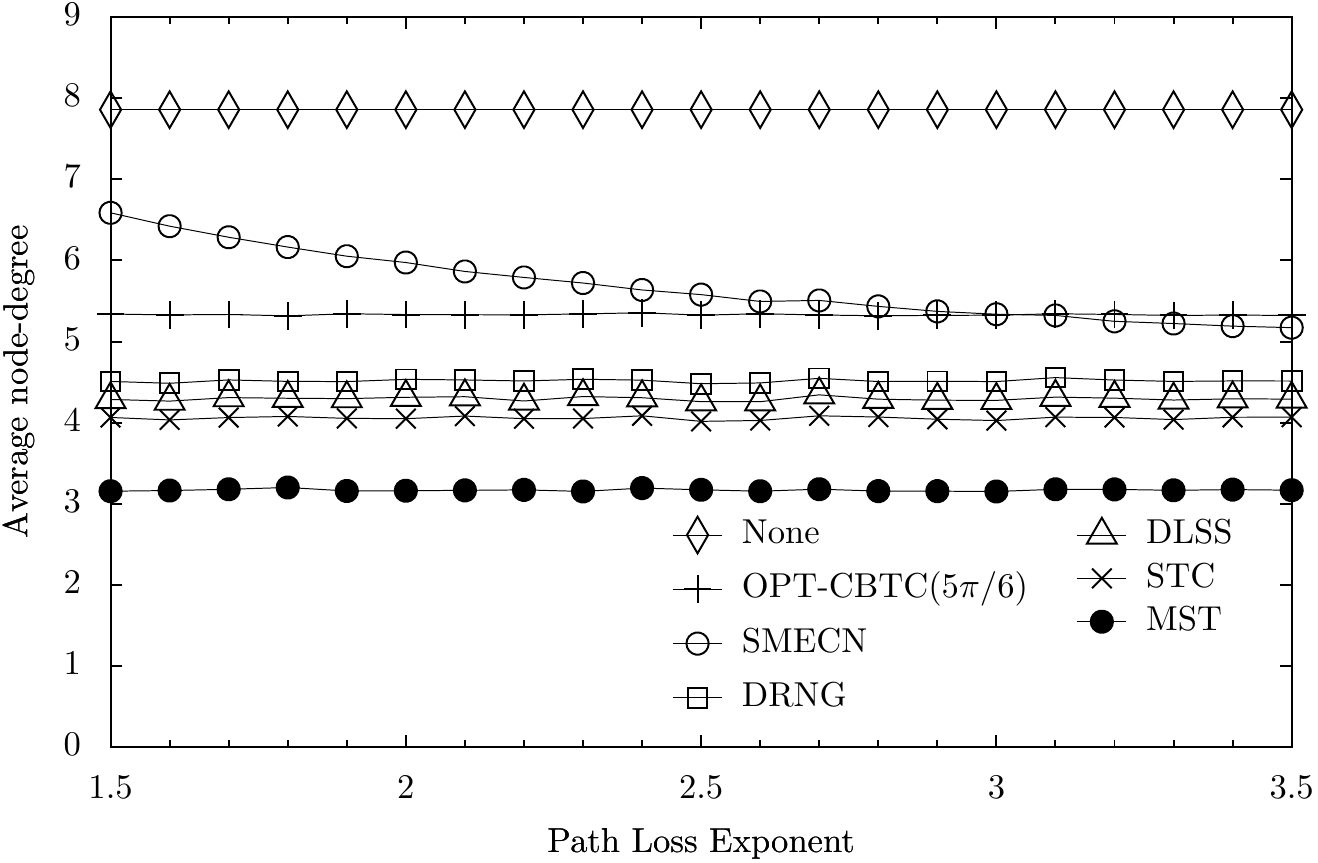}
        }
    \hskip0.1in 
    \subfigure[{$C_{T/H}(u \rightarrow v, {\mathrm Hops})$ averaged over all pairs of nodes $u$ and $v$.}]{
        \label{fig:Fig2_avgPathOmniPower}
        \includegraphics[width=3in]{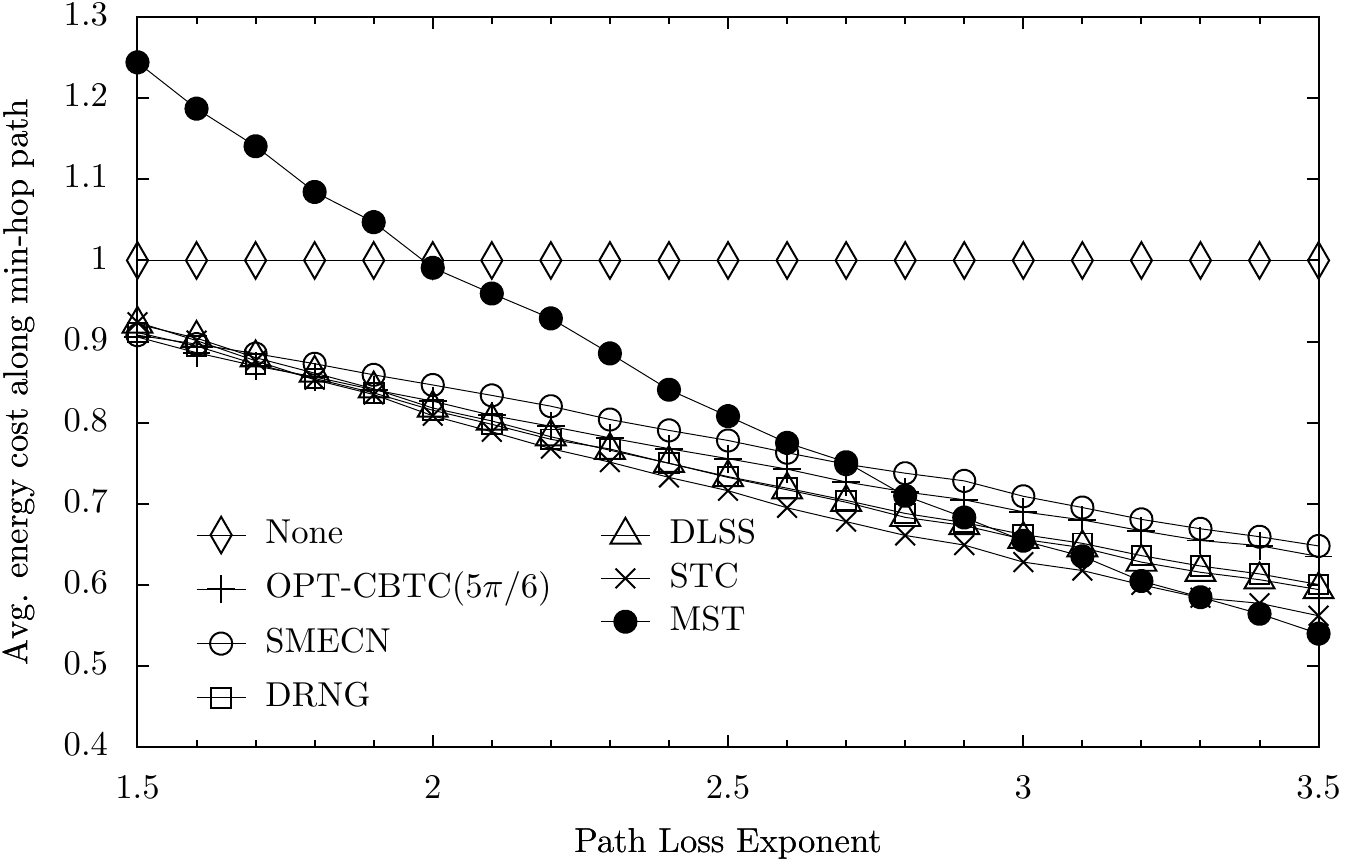}
        }\\
    \subfigure[{$I_{T/H}(u \rightarrow v, {\mathrm Hops})$ averaged over all
      pairs of nodes $u$ and $v$.}]{
        \label{fig:Fig2_avgPathOmniInterf}
        \includegraphics[width=3in]{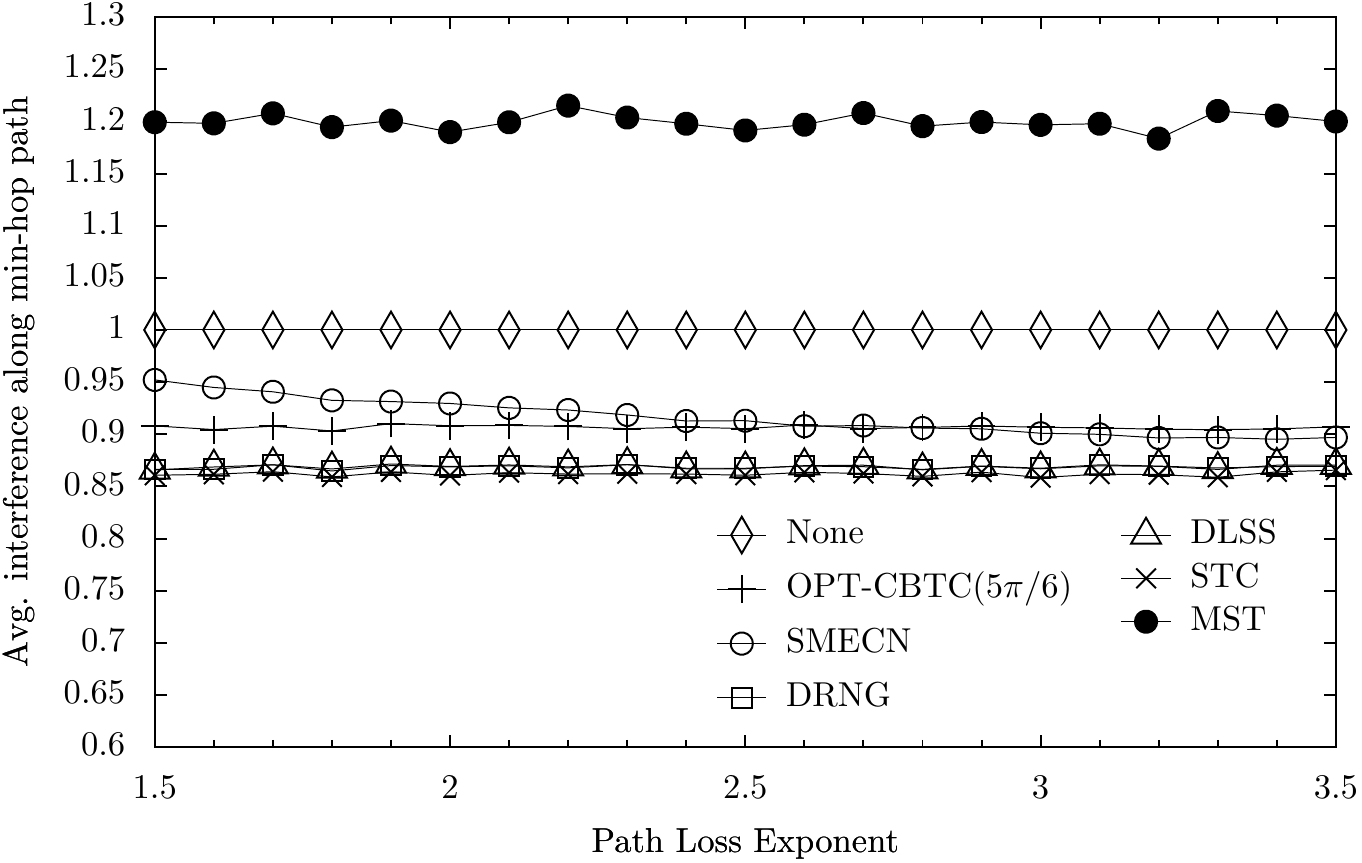}
        }
    \hskip0.1in 
    \subfigure[{$I_{T/H}(u \rightarrow v, {\mathrm Energy})$ averaged over all
      pairs of nodes $u$ and $v$.}]{
        \label{fig:Fig2_avgOptPathOmniInterf}
        \includegraphics[width=3in]{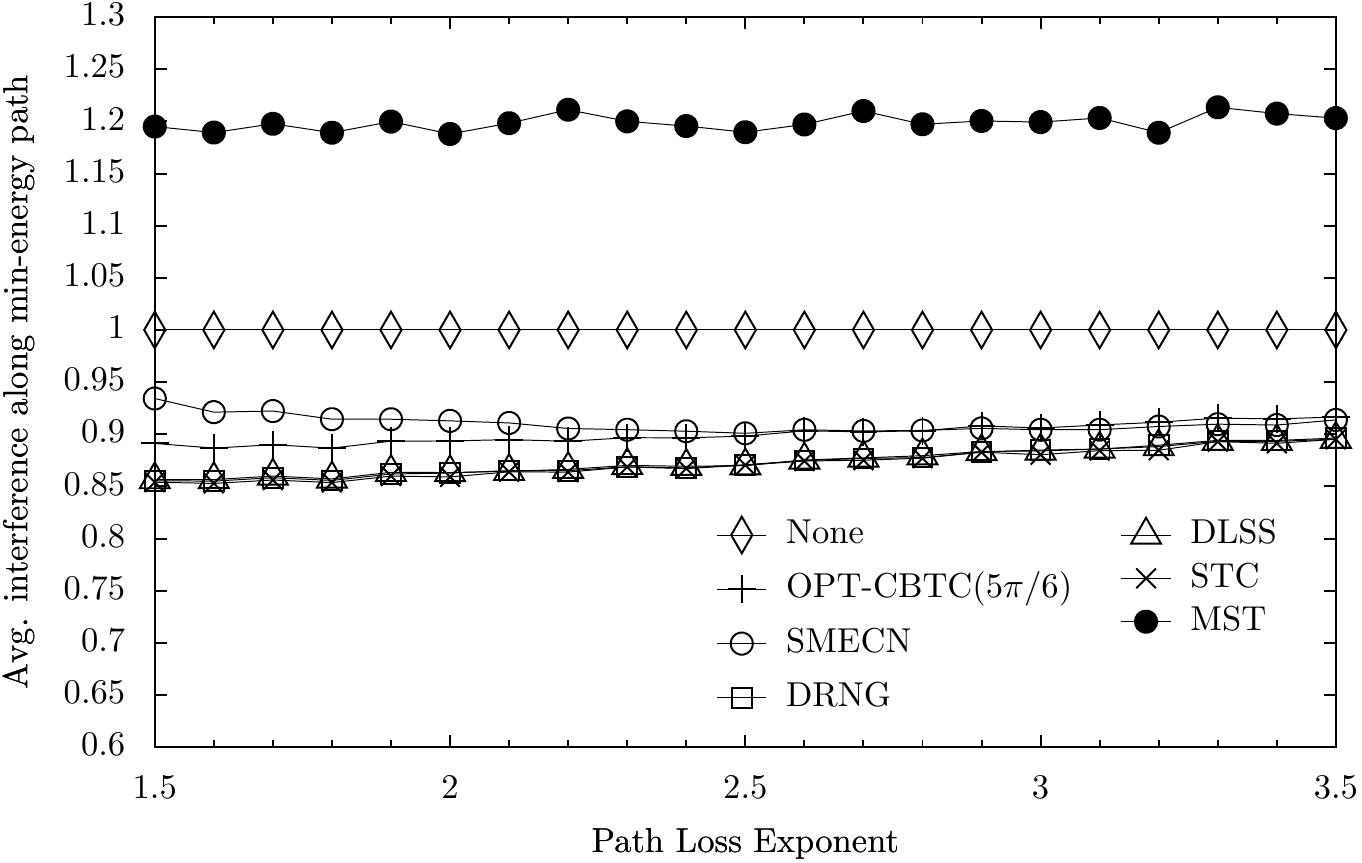}
        }
    \caption{Graphs showing the effectiveness of topology control
      algorithms when the path loss exponent is the same across the entire region
      of the network (in order to allow comparisons with the CBTC
      algorithms). The networks are generated with 200 randomly located 
      nodes in a unit square area. Each data point represents an
      average of one hundred random networks. Note that a longer
      distance between two nodes does not necessarily mean higher
      energy cost across the pair since path loss exponents follow a
      random Gaussian distribution.}
\end{center}
\end{figure*}

\begin{figure*}[!t]
\begin{center}
    \subfigure[{Average $P_{T/H}(u)$ over all $u$.}]{
        \label{fig:Fig3_avgOmniPower}
        \includegraphics[width=3in]{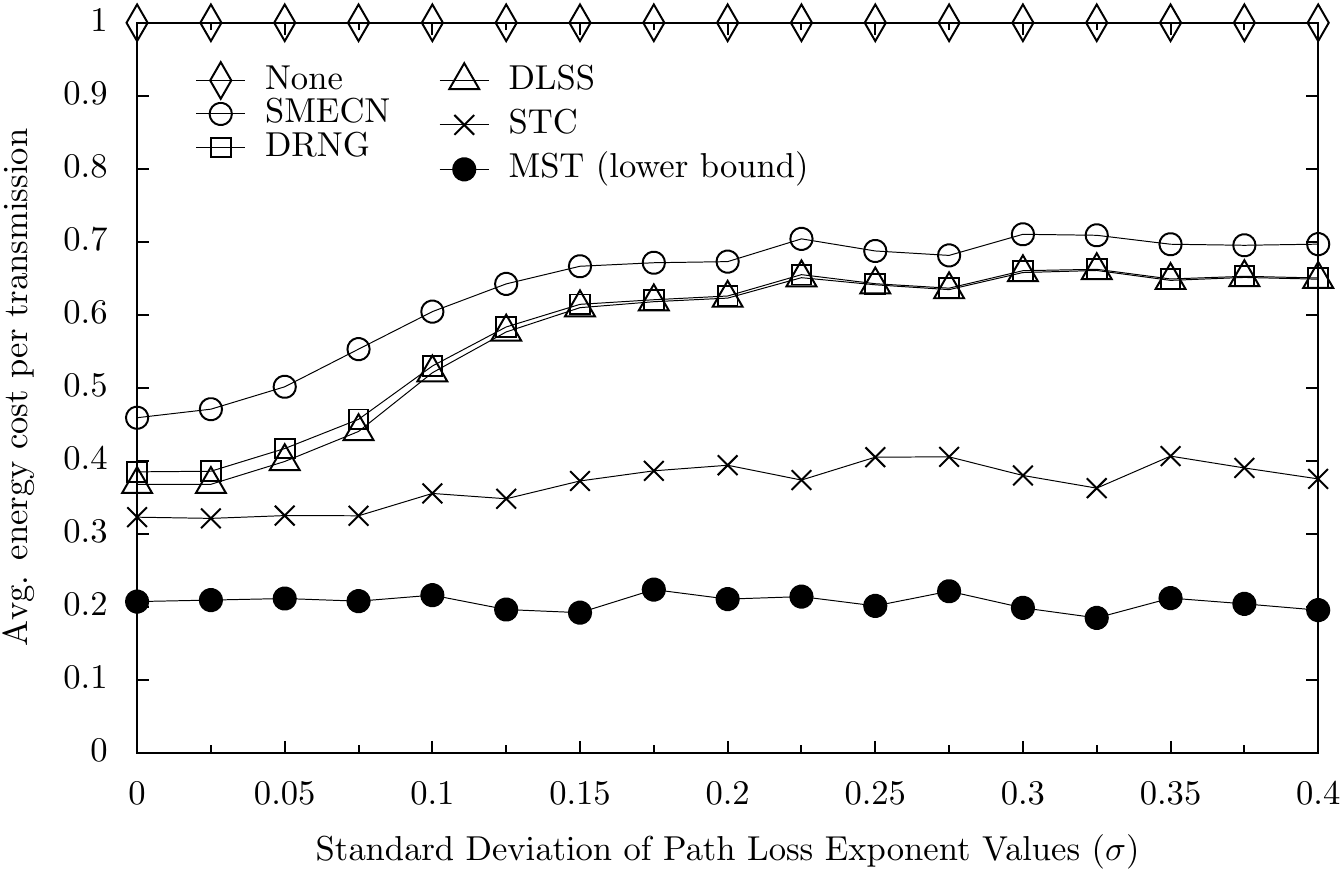}
        }
    \hskip0.1in 
    \subfigure[{$C_{T/H}(u \rightarrow v, {\mathrm Energy})$ averaged over all
      pairs of nodes $u$ and $v$.}]{
        \label{fig:Fig3_avgOptPathOmniPower}
        \includegraphics[width=3in]{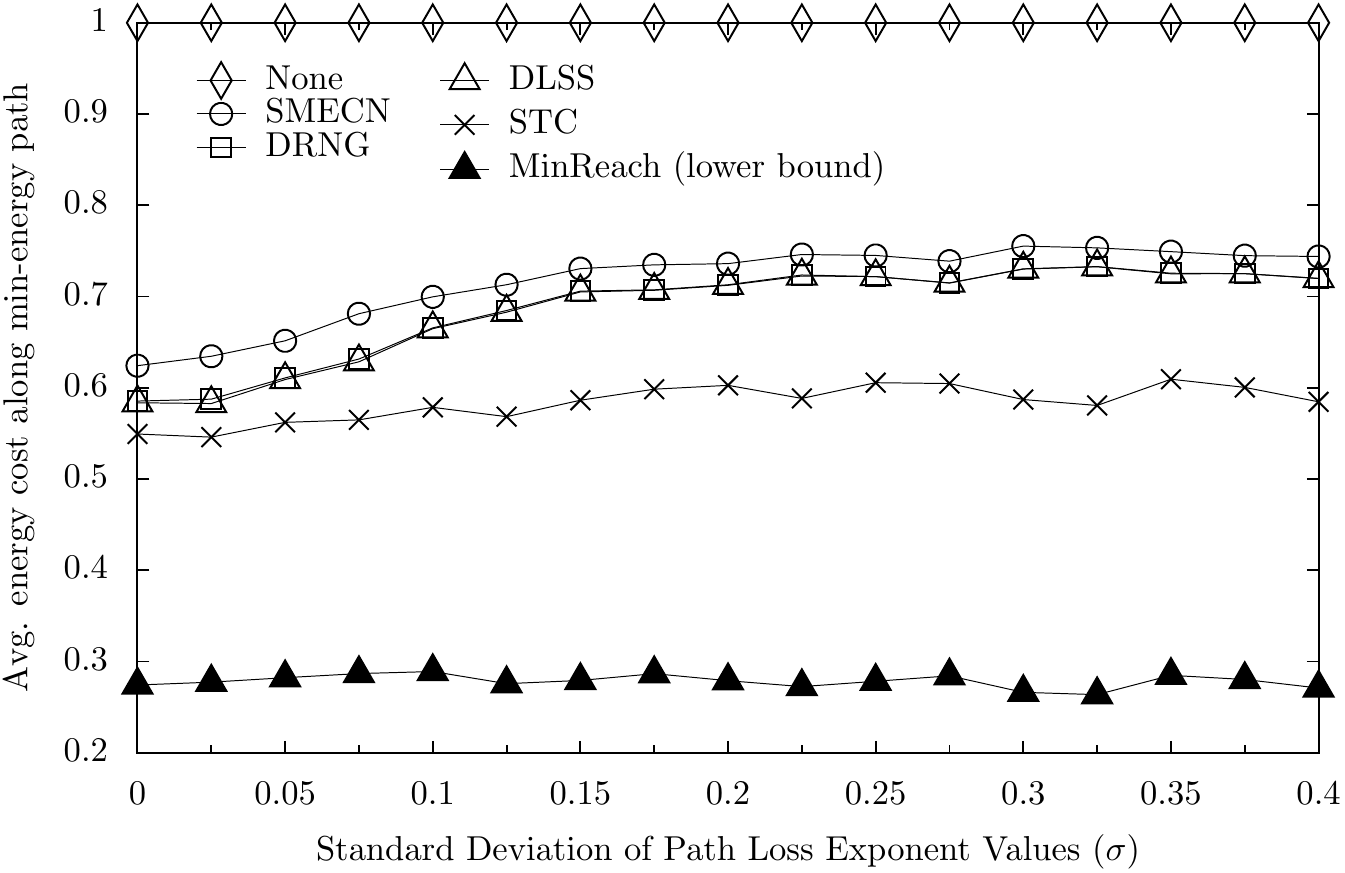}
        }\\
    \subfigure[{Average node-degree.}]{
        \label{fig:Fig3_avgOmniNodeDegree}
        \includegraphics[width=3in]{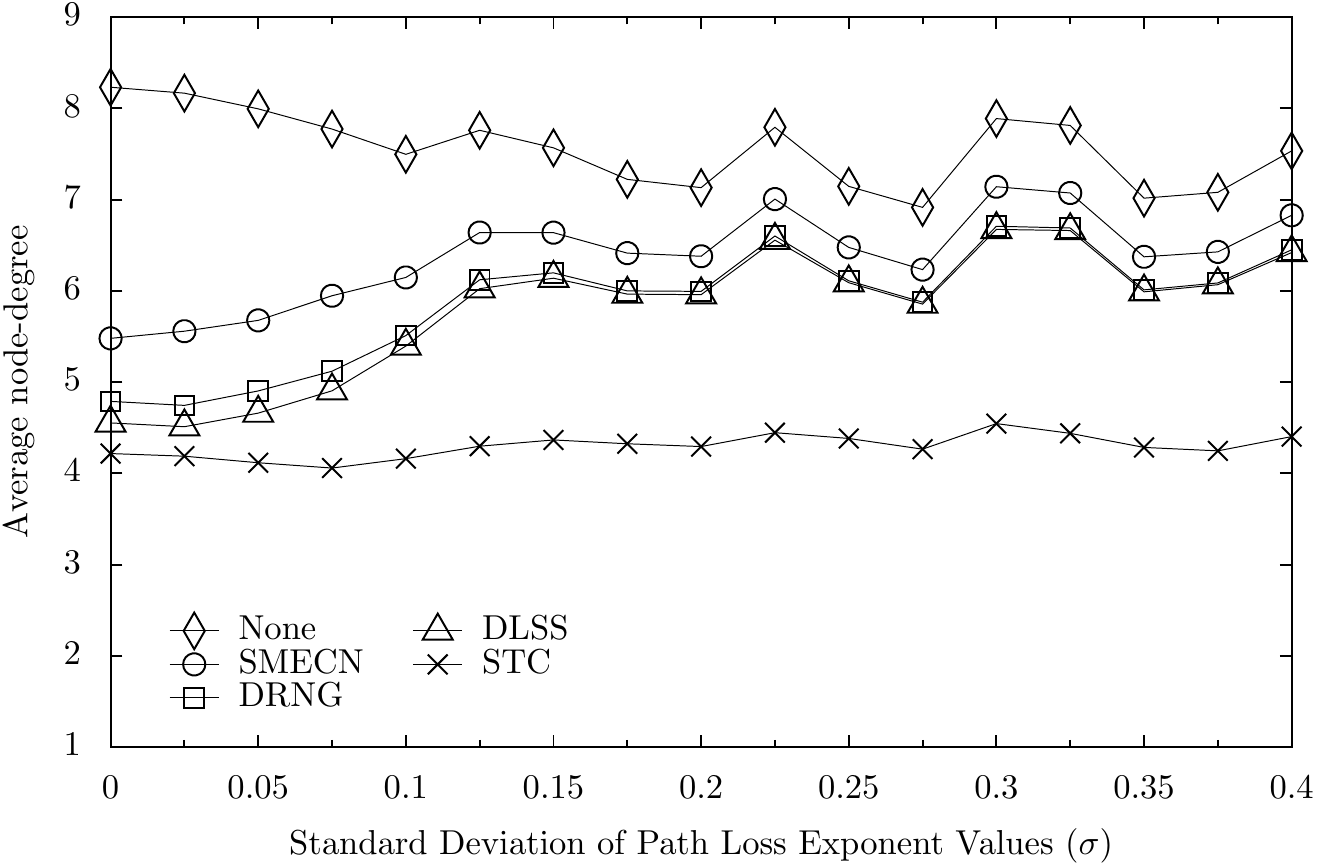}
        }
    \hskip0.1in 
    \subfigure[{$C_{T/H}(u \rightarrow v, {\mathrm Hops})$ averaged over all
      pairs of nodes $u$ and $v$.}]{
        \label{fig:Fig3_avgPathOmniPower}
        \includegraphics[width=3in]{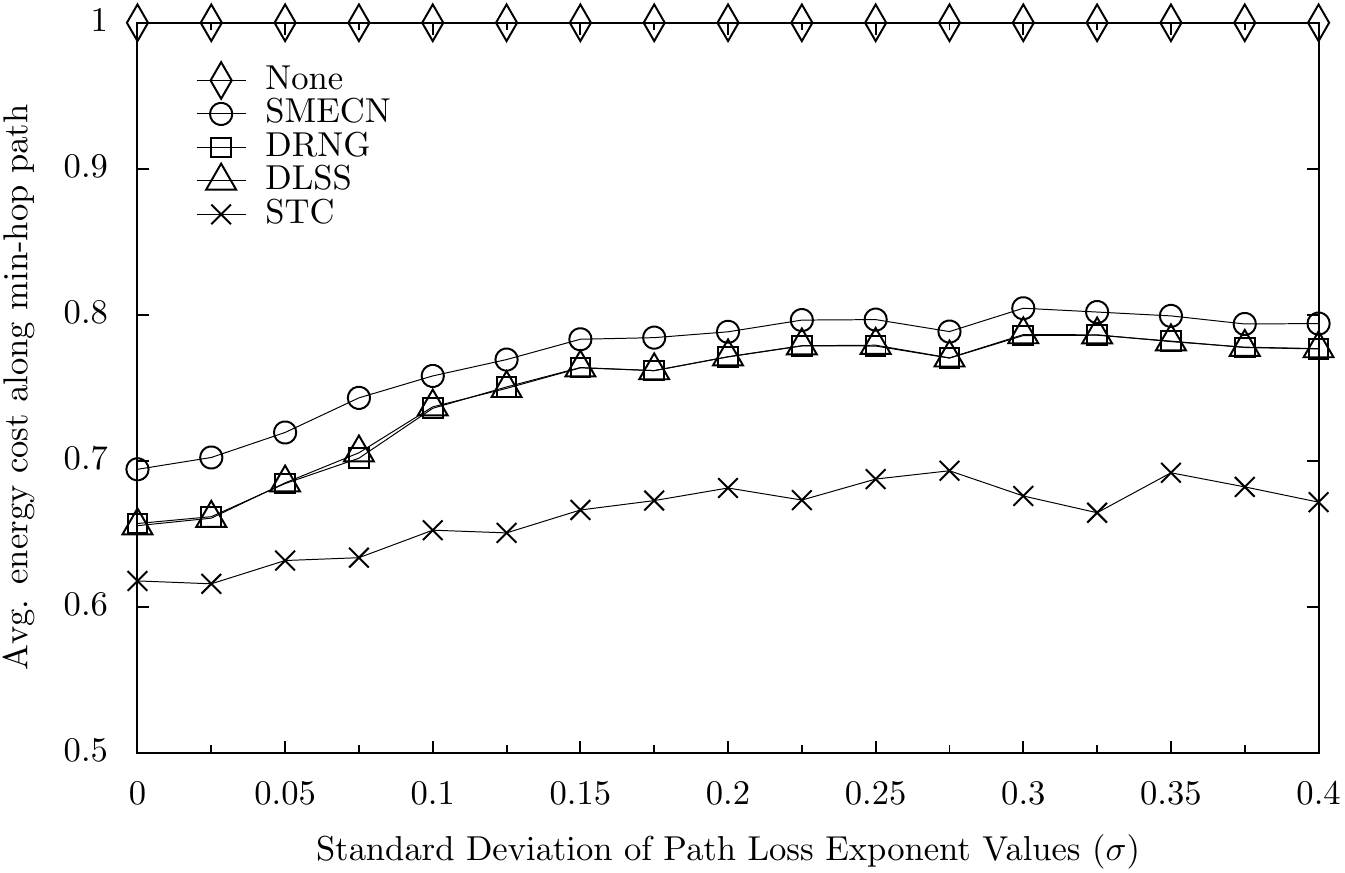}
        }\\
    \subfigure[{$I_{T/H}(u \rightarrow v, {\mathrm Hops})$ averaged over all
      pairs of nodes $u$ and $v$.}]{
        \label{fig:Fig3_avgPathOmniInterf}
        \includegraphics[width=3in]{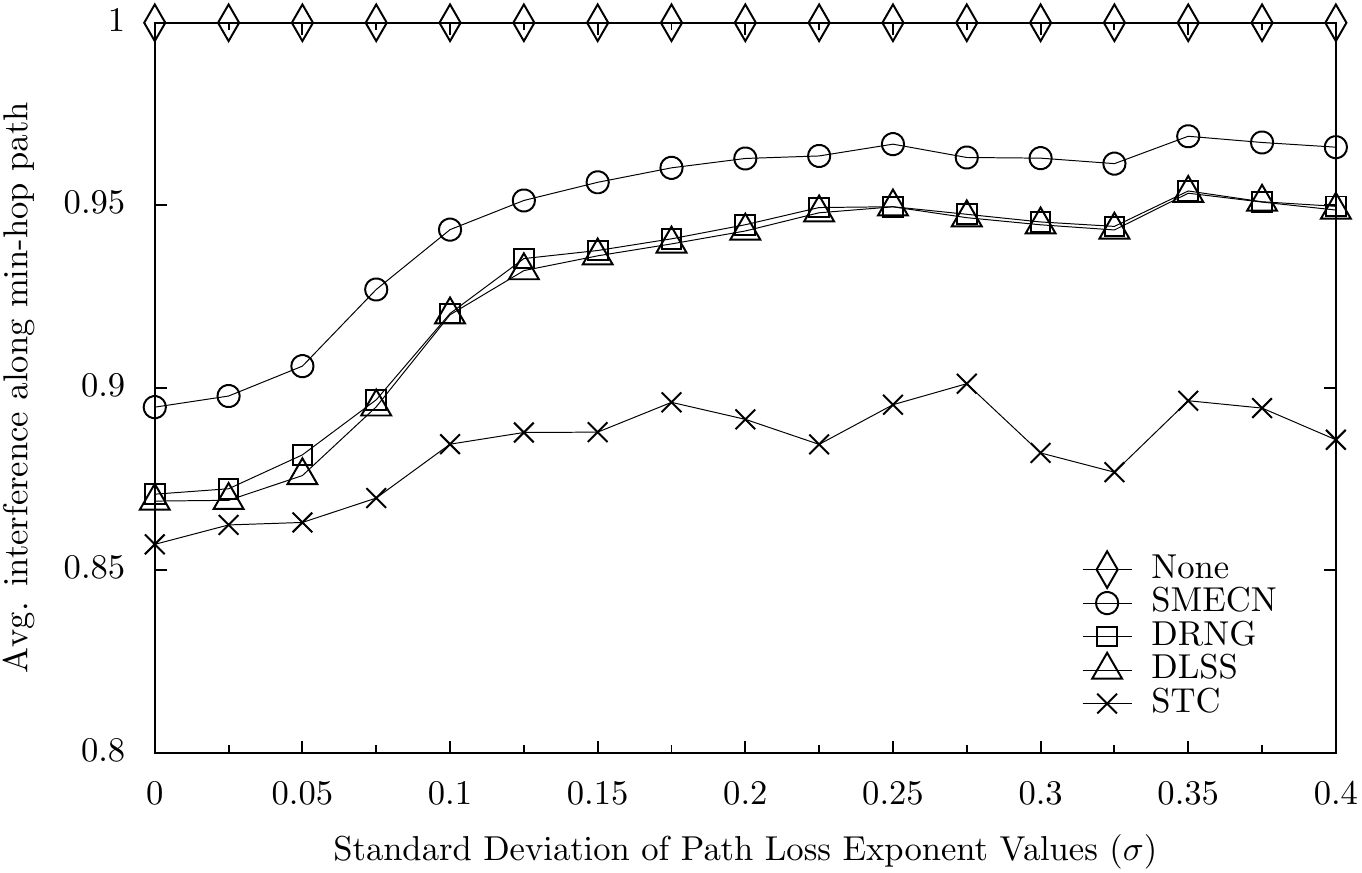}
        }
    \hskip0.1in 
    \subfigure[{$I_{T/H}(u \rightarrow v, {\mathrm Energy})$ averaged over all
      pairs of nodes $u$ and $v$.}]{
        \label{fig:Fig3_avgOptPathOmniInterf}
        \includegraphics[width=3in]{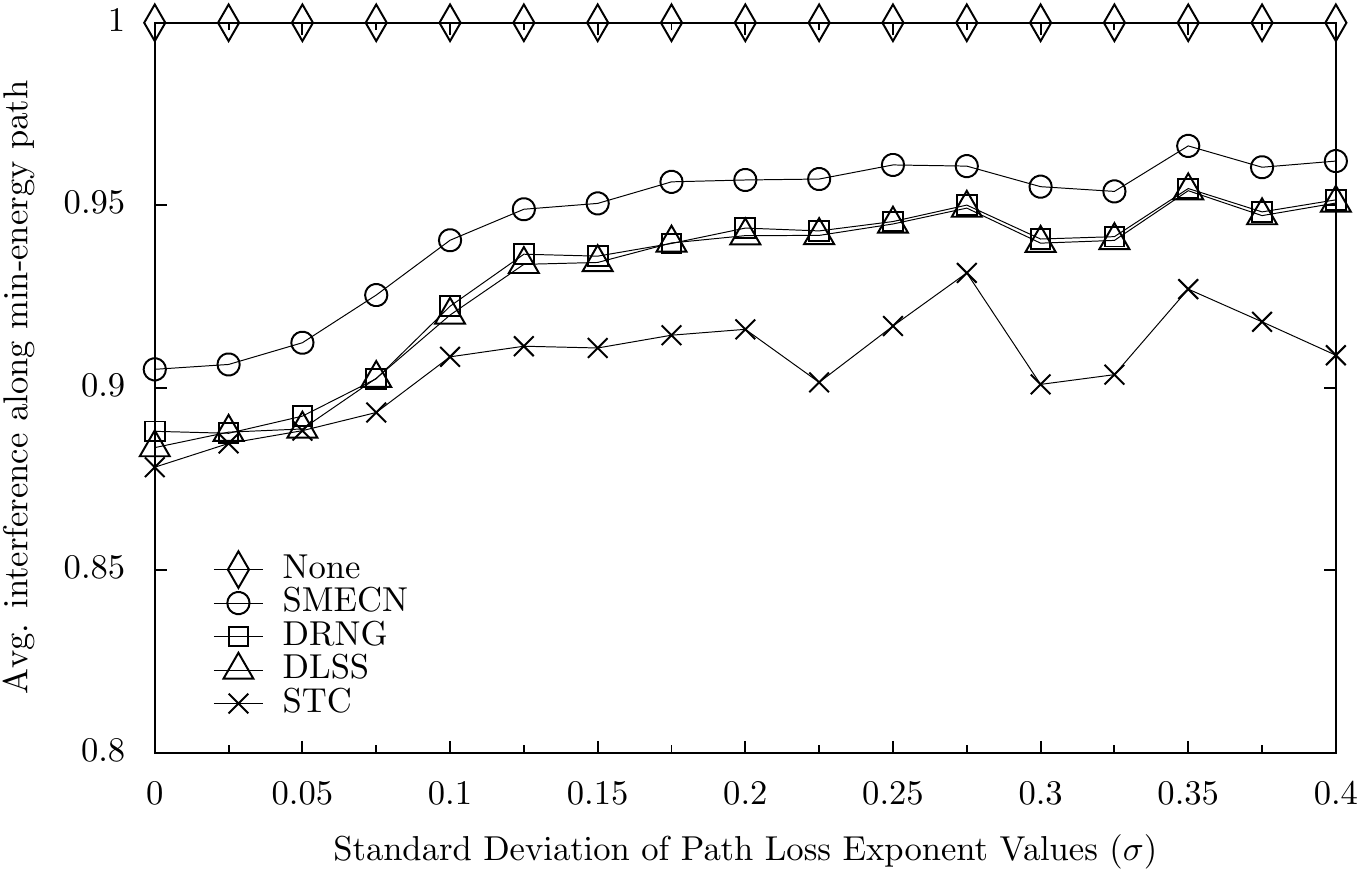}
        }
    \caption{Graphs showing the effectiveness of topology control
      algorithms in reducing energy costs when the path loss exponents
      exhibit a Gaussian distribution with a mean of 3.1 in the range
      $[2.7, 3.5]$ for various values of the standard deviation
      (indoor propagation \cite{Rap2002}). The networks are generated
      with 200 randomly located nodes in a unit square area.} 
\end{center}
\end{figure*}

\begin{figure*}[!t]
\begin{center}
    \subfigure[{Average $P_{T/H}(u)$ over all $u$.}]{
        \label{fig:Fig4_avgOmniPower}
        \includegraphics[width=3in]{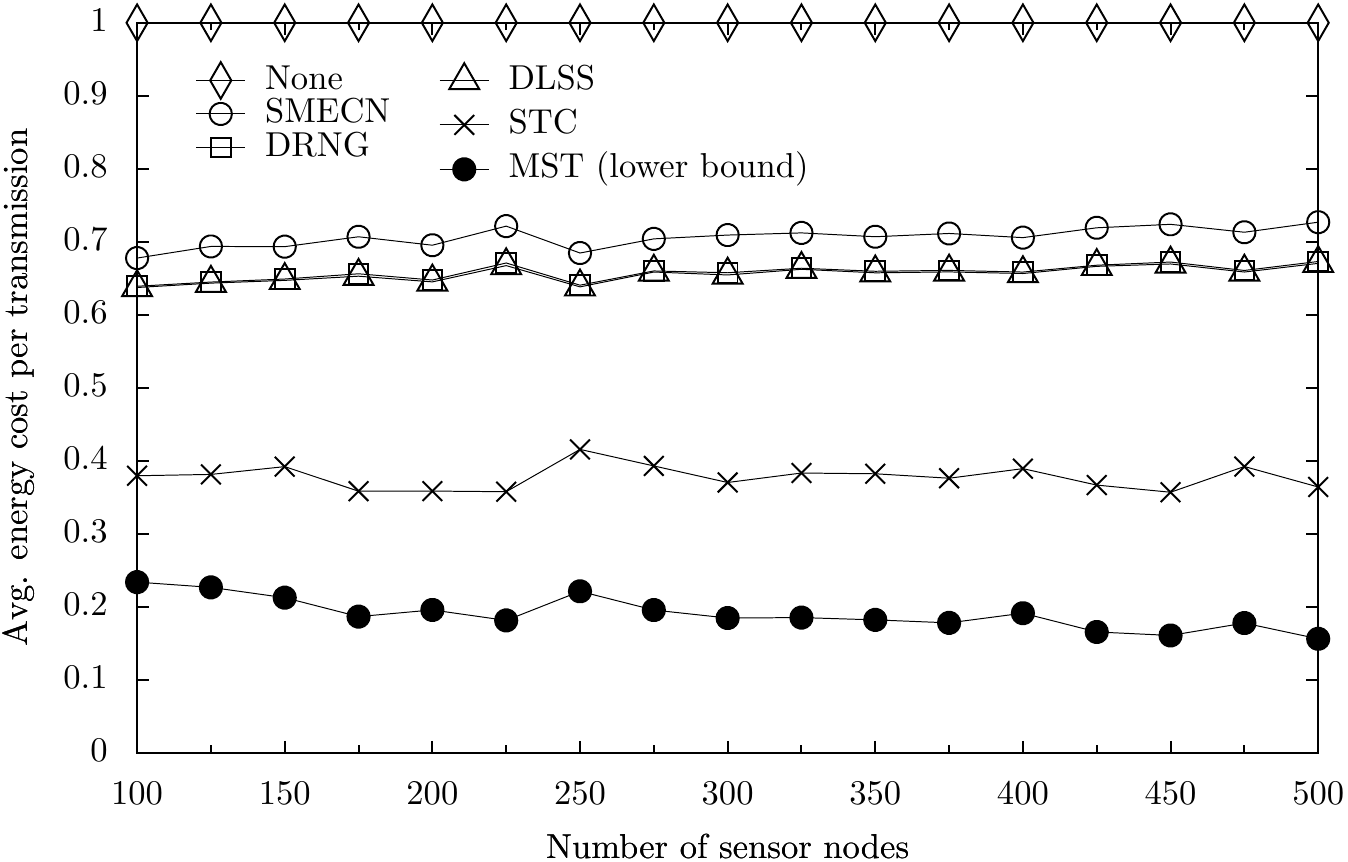}
        }
    \hskip0.2in 
    \subfigure[{$C_{T/H}(u \rightarrow v, {\mathrm Energy})$ averaged over all
      pairs of nodes $u$ and $v$.}]{
        \label{fig:Fig4_avgOptPathOmniPower}
        \includegraphics[width=3in]{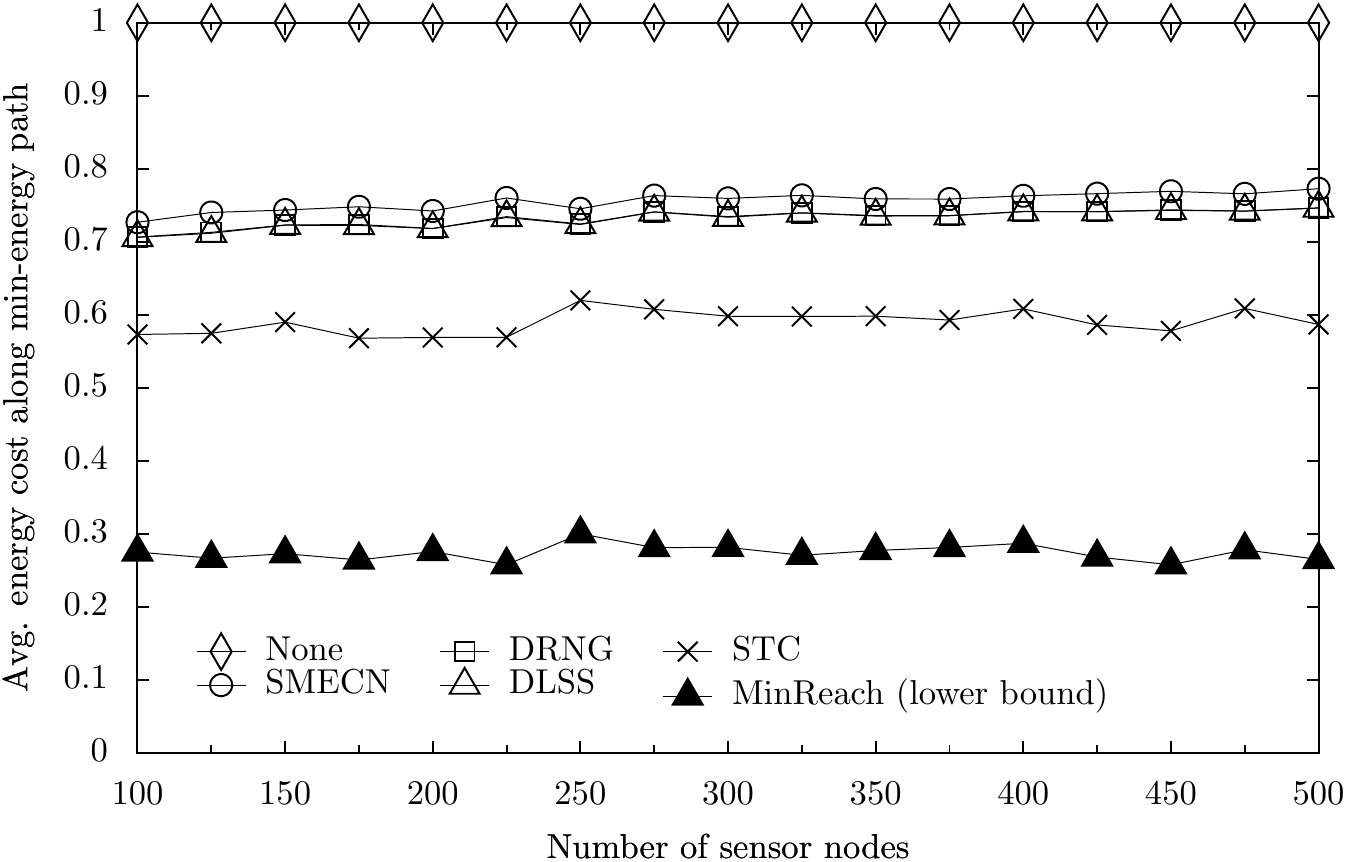}
        }
    \caption{Graphs showing the effectiveness of topology control
      algorithms in reducing energy costs when the path loss exponent
      varies around a mean of 3.1 with a standard deviation of 0.16
      (indoor propagation \cite{Rap2002}). The graphs show that the
      effectiveness of the STC algorithm (for that matter, all
      localized topology control algorithms) do not change with
      increase in the number of nodes.}
\end{center}
\end{figure*}

Our measure of interference derives from the definition in
\cite{JohCar2005}, which is a refinement of that used in
\cite{BurRic2004}. Define the span, span$(e)$, of an edge $e=(g,h)$ as the
number of nodes that are neighbors of at least one of $g$ and
$h$. This represents the number of nodes that would have to remain
silent to enable a successful transmission between $g$ and $h$. We measure
interference, $I_T(u  \rightarrow v, {\mathrm Hops})$, along the
minimum-hop path from $u$ to $v$
in cover graph $T$ as the sum of
span$(e)$ for all $e$ in the path. $I_{T/H}(u
\rightarrow v, {\mathrm Hops})$ denotes the ratio of the interference
along the minimum-hop path
from $u$ to $v$ in cover graph $T$ generated by the topology control
algorithm and the corresponding cost along the minimum-hop path
from $u$ to $v$ in the initial graph, $H$. This ratio
represents the reduction in interference achieved due to the topology
control algorithm. The ratio for the minimum-energy paths is computed
similarly as above and is denoted by $I_{T/H}(u
\rightarrow v, {\mathrm Energy})$.

\subsection{Wireless Model}
\label{wireless_model}

The pertinent issue in the modeling of the wireless environment in the
context of this paper is the path loss model. The log-distance path
loss model based on the path loss as a logarithmic function of the
distance $d$ has been confirmed both theoretically and by measurements
in a large variety of environments \cite{Rap2002}. In this
model, the path loss at distance $d$, $PL(d)$ is expressed as: 
\[
PL(d) = PL(d_0) + 10 \gamma \log_{10}( d/d_0 )
\]
where the constant $d_0$ is an arbitrary reference distance and
$\gamma$ is called the path loss exponent. 

The path loss exponent is an important parameter in modeling a
wireless environment and varies within a region depending upon a
number of factors including antenna characteristics, transmission
frequency, the nature of the obstructions, multipath, and shadowing
effects. While most of these effects are location-specific and
difficult to generalize across different environments
\cite{HayMoh2005}, empirical observations in several real environments
do reveal that the distribution of path loss exponents within a region
of interest is Gaussian
\cite{ErcGre1999,DabHai2006,LieRei2007,GenVai2007}. Our simulation
models, accordingly, use a Gaussian distribution when the path loss
exponent is assumed to vary across the region of interest.

In our simulation studies, the specific values of the path loss
exponents and the standard deviation of their distribution are based
on results from empirical studies of indoor propagation described in
\cite{Rap2002}. We choose the indoor propagation model because the 
constraints on topology control specifically considered in this paper,
such as having to avoid a GPS receiver and not being able to discern
the direction of signal arrival, are most applicable to indoor
wireless environments. Our study uses three sets of simulation
experiments in comparing the effectiveness of different topology
control algorithms.

In the first set of experiments, we assume that the path loss exponent
is the same across the entire network and study the topology control
algorithms for different values of the path loss exponent between 1.5
to 3.5. These experiments are intended to reveal the dependence of 
topology control performance on the path loss exponent and to allow
a comparison to CBTC algorithms (since they assume a uniform path loss
exponent across the entire region). This first set of experiments is
not intended to simulate a real wireless environment. 

In the second set of experiments, we vary the path loss exponent
randomly using a truncated Gaussian distribution, as suggested in
\cite{ErcGre1999} for simulation experiments. Based on empirical
measurements of indoor propagation at 634 locations in buildings
\cite{Rap2002}, the path loss exponents in our experiment follow a
truncated Gaussian distribution between 2.7 and 3.5 with a mean of
3.1. We vary the standard deviation of the distribution from 0 to 0.4
to study the impact of the spatial variation of path loss exponents on
the performance of topology control algorithms. 

In the third set of experiments, we study the effectiveness
of the algorithms as the number of nodes increases. The
networks used in our first two sets of experiments consist of 200
nodes located randomly in a unit square area. In the third set of
experiments, we vary the number of nodes from 100 to 500 to study the
scalability of the algorithms across a five-fold increase in the
number of nodes. In these experiments also, we use a truncated
Gaussian distribution for the path loss exponents ranging from 2.7 to
3.5 with a mean of 3.1. However, we use a single standard deviation of
0.16 (based on scatter plots from the empirical measurements of indoor
propagation on a single floor of a building, described in
\cite{Rap2002}). Each data point in this paper represents an average
of one hundred different randomly generated networks.

Finally, we note that DRNG and STC
algorithms can be generalized by a parameter $k$ so that an edge
$(u,v)$ is removed when there exists a path of $k$ hops or less
from $u$ to $v$ and from $v$ to $u$ provided the energy cost across
each of these hops is less than that corresponding to a direct
transmission between $u$ and $v$. In the fourth set of experiments,
we examine the dependence of our key performance metrics on $k$.

Our simulation model, because it is based on empirical measurements in
real wireless environments, encapsulates multipath and other
phenomena such as shadowing common in real environments. The following
describes each of the simulation experiments and the
results in greater detail.  

\subsection{Experiment 1}

We include the CBTC algorithms in our first set of simulation experiments. The CBTC
algorithms, however, assume that the loss propagation characteristics
of the medium are uniform across the entire region. In fairness to the
CBTC algorithms, therefore, we conduct this simulation experiment with
identical path loss exponents between any pair of directly
communicating nodes in the network. 

For path loss exponents ranging from 1.5 to 3.5, we first provide
results on the two objectives of the topology control problem
statement described in \ref{problem_statement}. Figure
\ref{fig:Fig2_avgOmniPower} presents the ratio $P_{T/H}(u)$ averaged
over all nodes. The cover graph of the minimum spanning tree (MST)
represents the lower bound of the achievable average $P_{T/H}(u)$ and
is also plotted in the graph. Figure \ref{fig:Fig2_avgOptPathOmniPower} 
plots the ratio $C_{T/H}(u \rightarrow v, {\mathrm Energy})$ averaged
over all pairs of nodes $u$ and $v$. The result obtained for MinReach
represents the lower bound on the average energy costs along a
path. Note that the MST graph which achieves the lower bound for the
first objective does very poorly on the second objective of minimizing
the energy cost along paths. This is because the MST graph has far too
few edges leading to unnecessarily long multi-hop paths. In fact, the
MST graph is worse than no topology control at all for low values of
path loss exponents.

Figure \ref{fig:Fig2_avgOmniNodeDegree} plots the 
average node-degree of a node in the cover graphs generated by the
topology control algorithms. SMECN is the only algorithm for which the
average node degree varies substantially with a change in the path
loss exponent. This is because SMECN makes a comparison between the
energy cost along an edge with the {\em sum} of the energy costs along
other edges. The result of such a comparison changes with the assumed
value of the path loss exponent and therefore, SMECN results in
different topologies under different values of the path loss
exponent. All other algorithms are based on {\em one-to-one}
comparison of edges as regards their energy costs, and the result of
such comparisons is the same independent of the path loss exponent as
long as the same path loss exponent characterizes the entire network.

Figure \ref{fig:Fig2_avgPathOmniPower} plots the ratio $C_{T/H}(u
\rightarrow v, {\mathrm Hops})$ averaged over all pairs of nodes $u$ and
$v$  along minimum-hop paths. This represents the energy costs along a
path when using routing algorithms that minimize the hop count. 
Figures \ref{fig:Fig2_avgPathOmniInterf} and
\ref{fig:Fig2_avgOptPathOmniInterf} similarly present results on
the average interference encountered in a path. 

These
results show that STC performs better---though only slightly
better---than the other topology control algorithms when the path loss
exponents are constant across the network. These results also show that
even while the STC algorithm generates a sparser graph than other existing
algorithms, it manages to keep the paths from lengthening
unnecessarily and thus achieves an overall reduction in energy
consumption. The same cannot be said of the MST graph, which performs
very poorly as regards interference. The MST graph has so few edges in
comparison to the original graph that paths between node pairs become
significantly longer, causing even more interference than one would
have with no topology control at all. 

Based on this set of experiments, we do not consider MST or a
distributed version of it as a candidate for topology
control. However, it is a useful metric as a lower bound on the average
transmission power of each node after the execution of a topology
control algorithm. Therefore, in the simulation results that follow,
we plot the results for MST only when presenting the ratio $P_{T/H}(u)$ averaged
over all nodes (the first objective of topology control in this
paper).

\subsection{Experiment 2}

In our second set of experiments, we allow a non-uniform value of the
path loss exponent in the region of interest. For each pair of
directly communicating nodes, we choose a random value of the path
loss exponent assuming a truncated Gaussian distribution with a mean
of 3.1 and ranging between 2.7 and 3.5 (based on the empirical
measurements cited in Section~\ref{wireless_model}). We consider
standard deviations of the Gaussian distribution ranging from 0 to
0.4. Since the CBTC algorithms work under the assumption that the loss
propagation characteristics of the medium are uniform across the
entire region, we do not include CBTC algorithms in this set of
experiments. 

If the energy cost of transmission from $u$ to $v$ is not the same as
that from $v$ to $u$, under DRNG or DLSS it is sometimes
possible for a node $u$ to keep a directed edge to node $v$, but for node $v$
to drop the directed edge to $u$. However, since many MAC layer protocols
expect bidirectional communication between
directly communicating nodes \cite{IEEE802.11std}, a topology control algorithm should
ideally generate a graph in which a directed edge from $u$ to $v$
exists if and only if a directed edge from $v$ to $u$
exists. Therefore, in order to ensure a fair comparison when we
simulate DRNG or DLSS in our studies, as in Section
\ref{sec:Complexity}, we assume that the energy cost or the path loss
exponent between two directly communicating nodes is the same in both
directions because the number and type of obstructions along each ray
path is likely the same even in multipath environments
\cite{IraBel2006}. 

Figure \ref{fig:Fig3_avgOmniPower} plots the ratio $P_{T/H}(u)$
averaged over all nodes for different degrees of variation in path
loss exponents. It is of interest to note that even though the STC
algorithm does not improve performance by much when the path loss
exponents are uniform across the network, it does make a {\em
  significant} difference when the path loss exponents vary. 

Figure \ref{fig:Fig3_avgOptPathOmniPower} plots the ratio $C_{T/H}(u
\rightarrow v, {\mathrm Energy})$ averaged over all pairs of nodes $u$
and $v$. Figure \ref{fig:Fig3_avgOmniNodeDegree} plots the average node
degree in the cover graphs generated by the various topology control
algorithms. Figure \ref{fig:Fig3_avgPathOmniPower} plots the ratio $C_{T/H}(u
\rightarrow v, {\mathrm Hops})$ averaged over all pairs of nodes $u$
and $v$. Figures \ref{fig:Fig3_avgPathOmniInterf} and
\ref{fig:Fig3_avgOptPathOmniInterf} similarly plot the
average interference along the minimum-hop and the minimum-energy
paths, respectively. These plots again indicate that even though the STC 
algorithm generates a sparser graph, it does not result in paths that
are so much longer that the energy consumed or the interference along
a path actually increases. In fact, we find that the STC algorithm
reduces the energy cost and the interference along a
path significantly in comparison to other topology control algorithms
that can be employed in real urban environments. In particular, the
effectiveness of the STC algorithm in cutting down the interference as
well as the energy costs along a path improves with increase in the
variation in path loss exponents (especially for small values of the
standard deviation in path loss exponent distribution).

Since the DLSS algorithm is the closest in performance to the STC
algorithm, it is worthwhile discussing the reasons behind the
significant difference between their performances in the presence of
variation in path loss exponents. Even though both DLSS and STC
algorithms incur the same order of initial information exchange
overhead, only in the STC algorithm does a node $u$ use the 
information about a node that is not
directly reachable by $u$ but is a common neighbor of two or more neighbors
of $u$. In DLSS, the local subgraph at node $u$ used for generating a
localized spanning tree in DLSS is one that is induced by the
neighbors of $u$ and does not contain a node that is not directly
reachable by $u$. STC performs better than DLSS because, in an
irregular wireless environment, it is more likely that a node, say $n$, is
unreachable by a node $u$ even if it is reachable by more than one
neighbor of $u$. Thus, DLSS ignores node $n$ in the creation of the
local spanning tree while the STC algorithm will consider it as long
as $n$ is reachable by some neighbor of $u$. 

\subsection{Experiment 3}

\begin{figure*}[!t]
\begin{center}
    \subfigure[{Initial graph, $H$}]{
        \label{fig:maxPower}
        \includegraphics[width=1.4in]{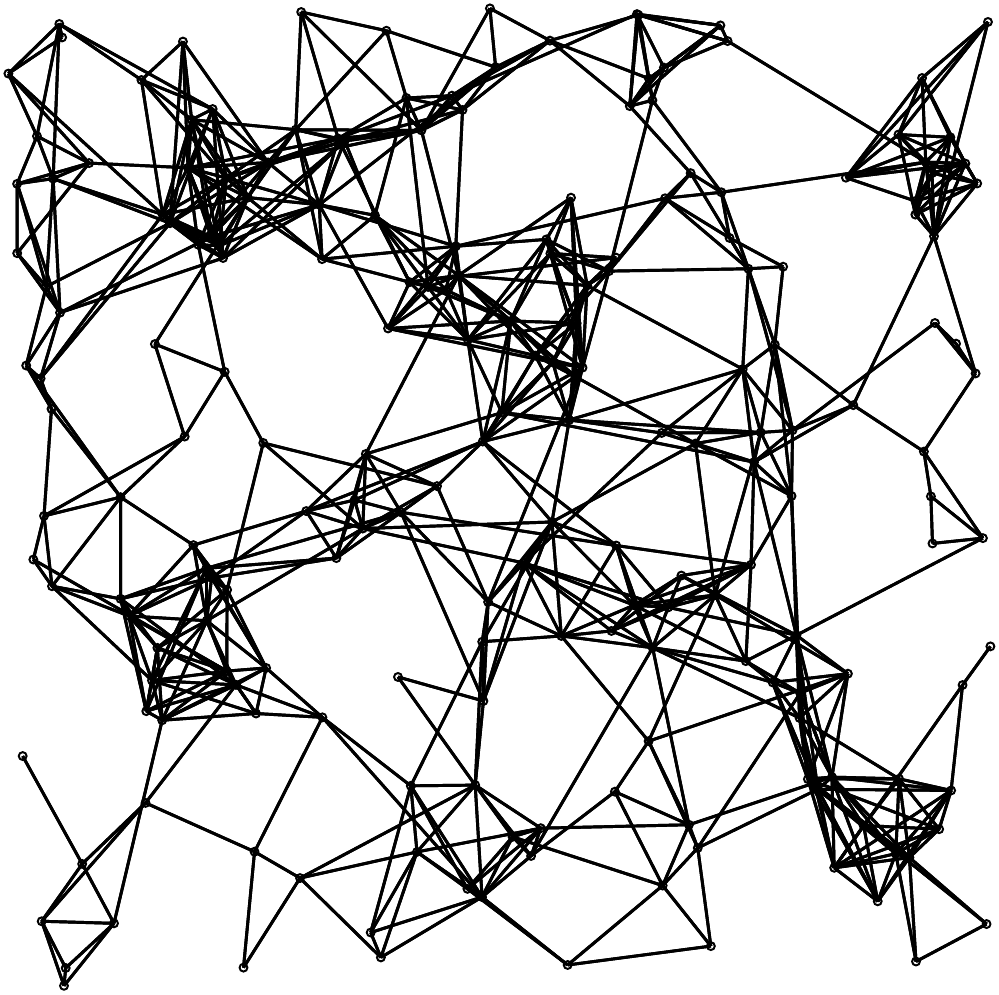}
        }
    \subfigure[{OPT-CBTC($5\pi/6$)}]{
        \label{fig:OPT_CBTC_5PIby6}
        \includegraphics[width=1.4in]{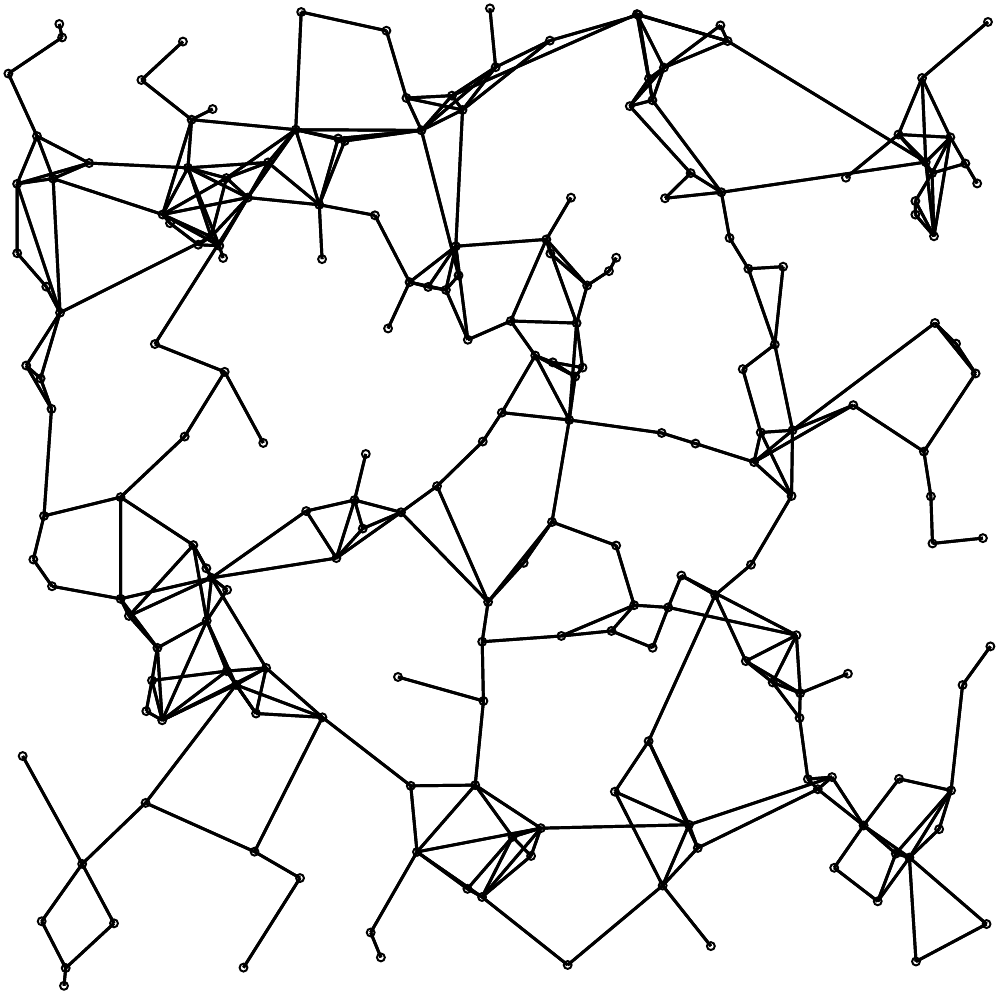}
        }
    \subfigure[{SMECN}]{
        \label{fig:SMECN}
        \includegraphics[width=1.4in]{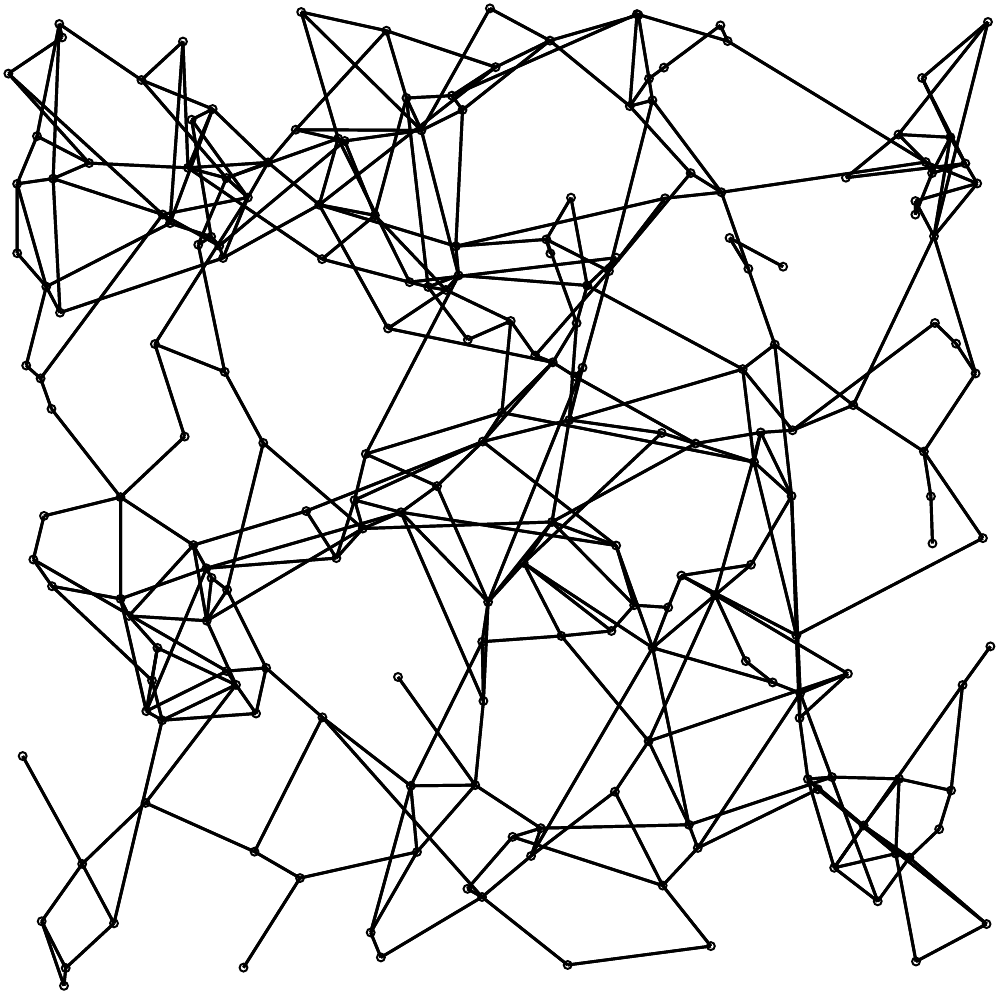}
        }
    \subfigure[{DRNG}]{
        \label{fig:DRNG}
        \includegraphics[width=1.4in]{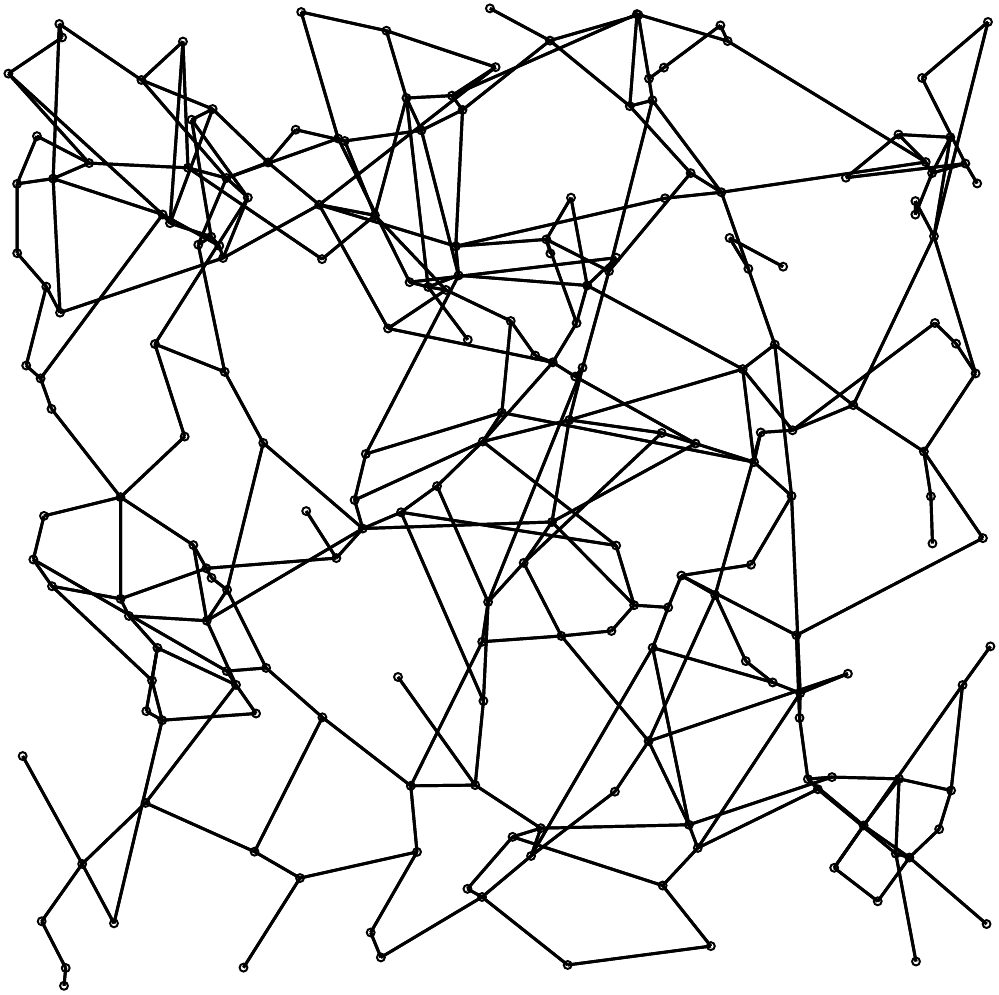}
        }\\
    \subfigure[{DLSS}]{
        \label{fig:DLSS}
        \includegraphics[width=1.4in]{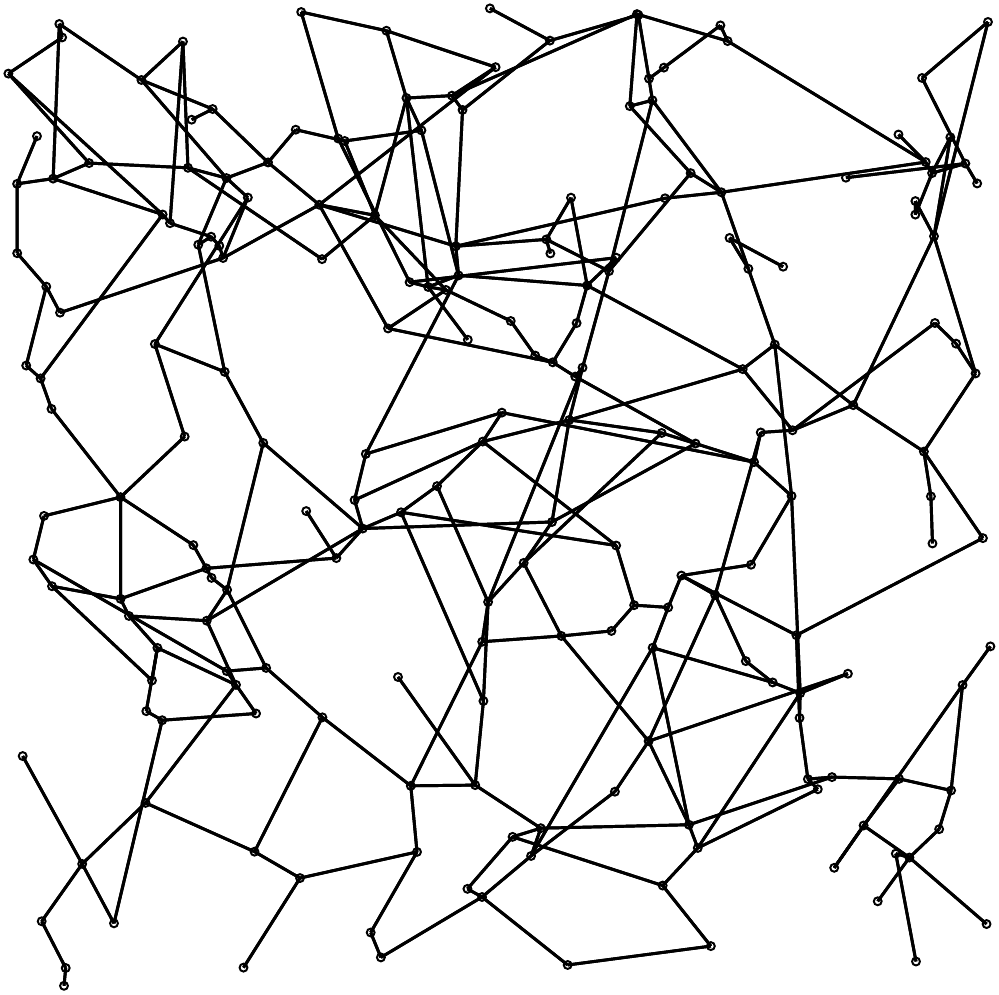}
        }
    \subfigure[{STC}]{
        \label{fig:STC}
        \includegraphics[width=1.4in]{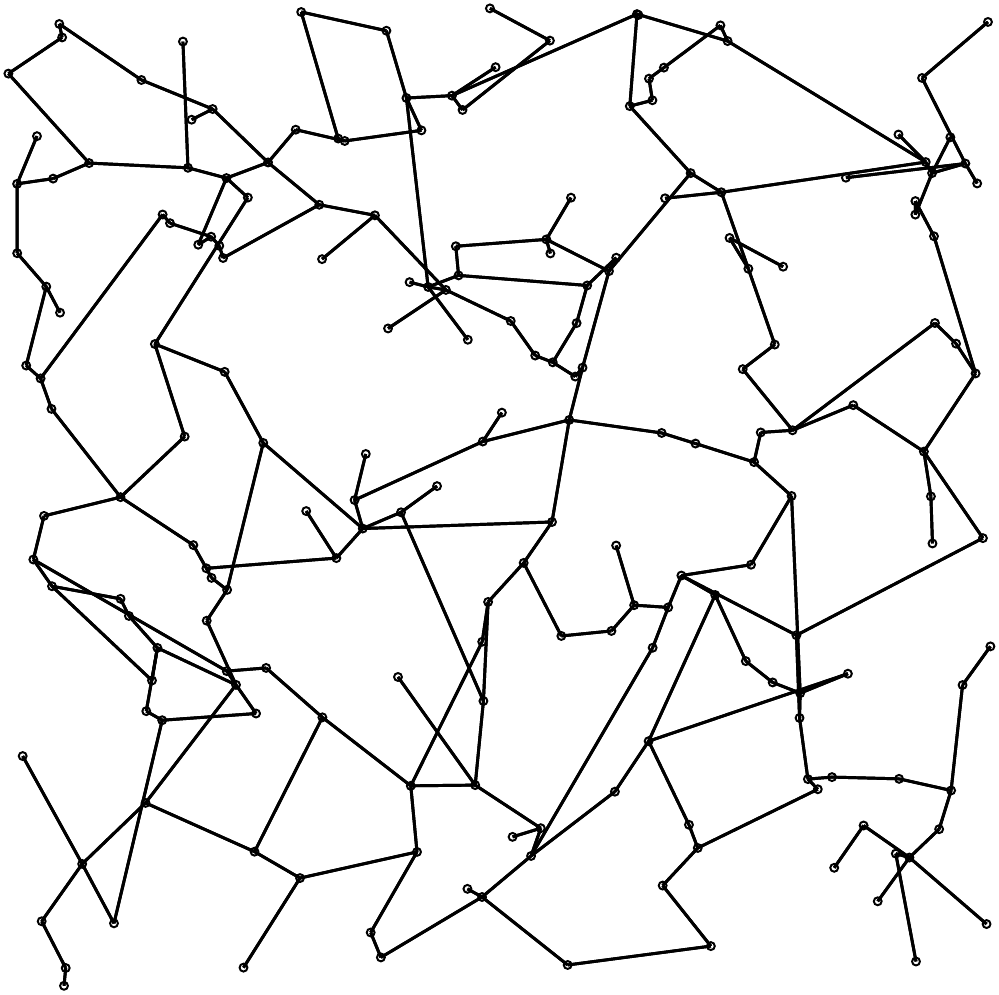}
        }
    \subfigure[{MST}]{
        \label{fig:MST}
        \includegraphics[width=1.4in]{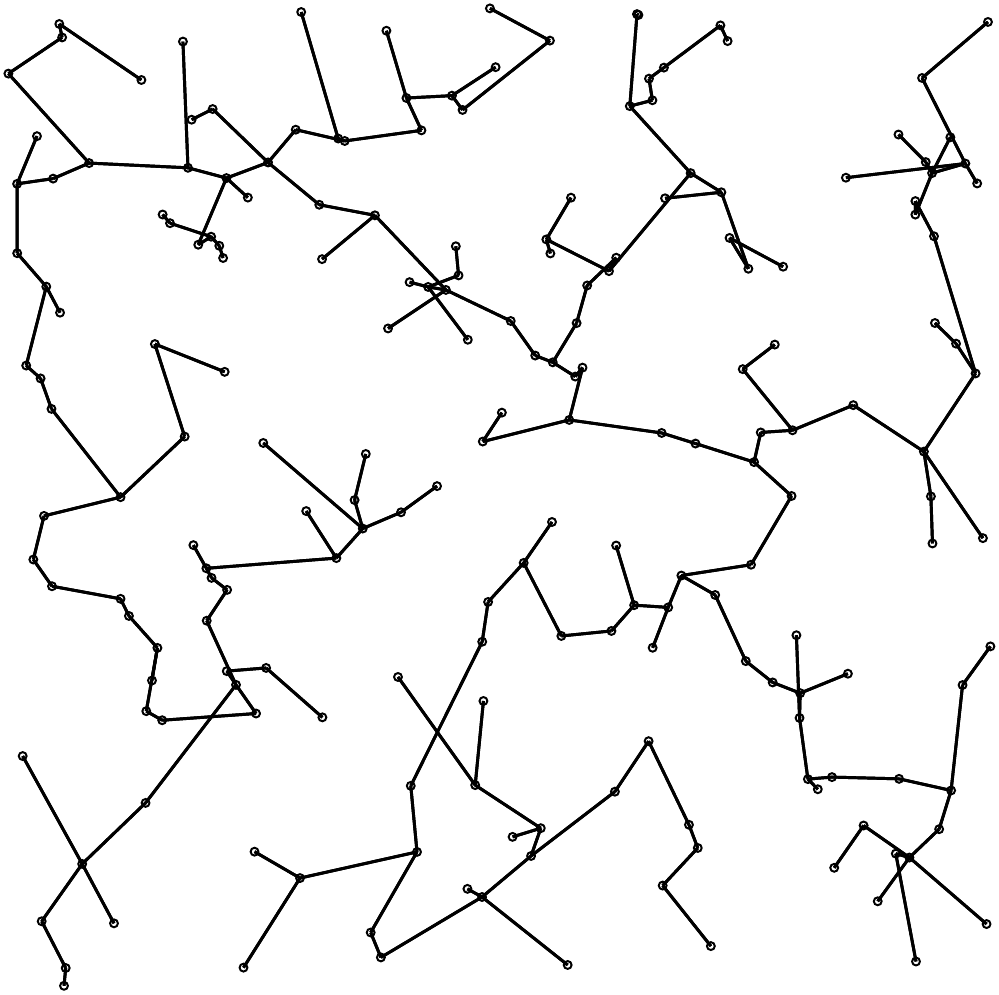}
        }\\

    \subfigure[{DLSS (cover graph)}]{
        \label{fig:DLSS_Omni}
        \includegraphics[width=1.4in]{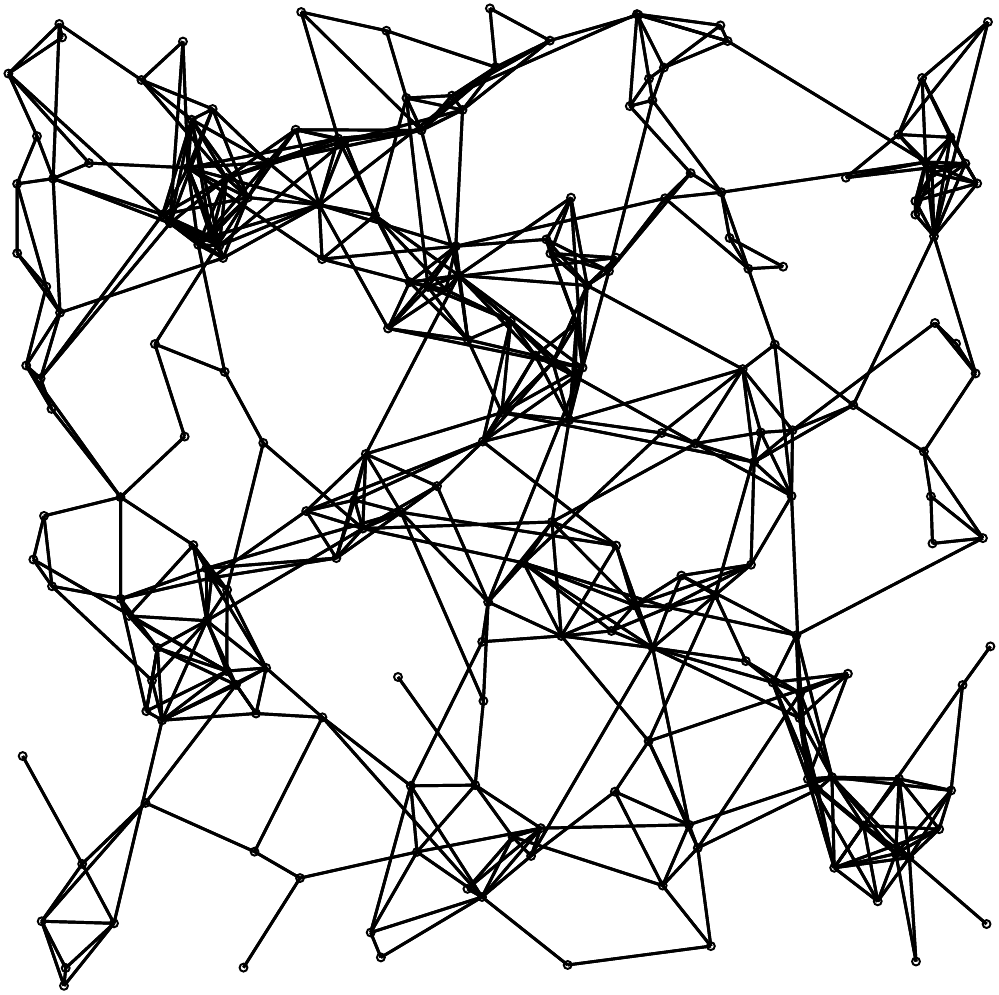}
        }
    \subfigure[{STC (cover graph)}]{
        \label{fig:STC_Omni}
        \includegraphics[width=1.4in]{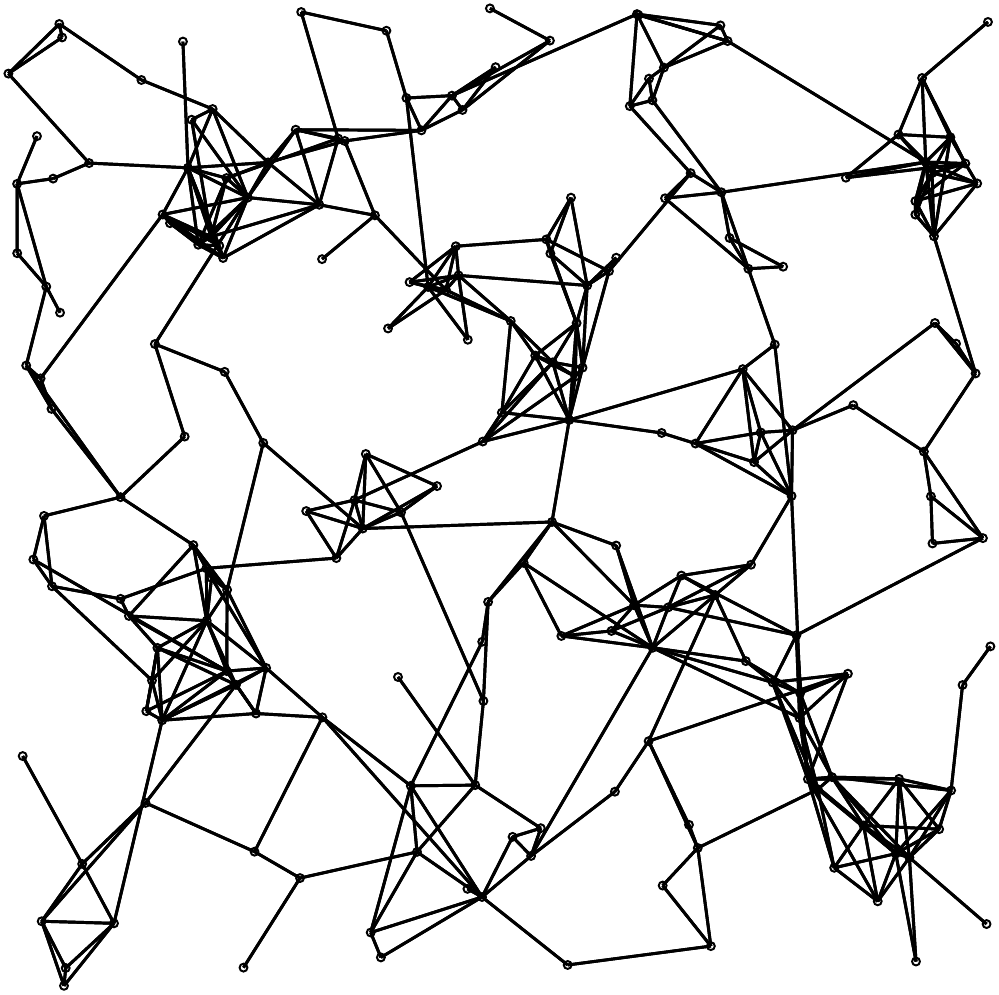}
        }
    \subfigure[{MST (cover graph)}]{
        \label{fig:MST_Omni}
        \includegraphics[width=1.4in]{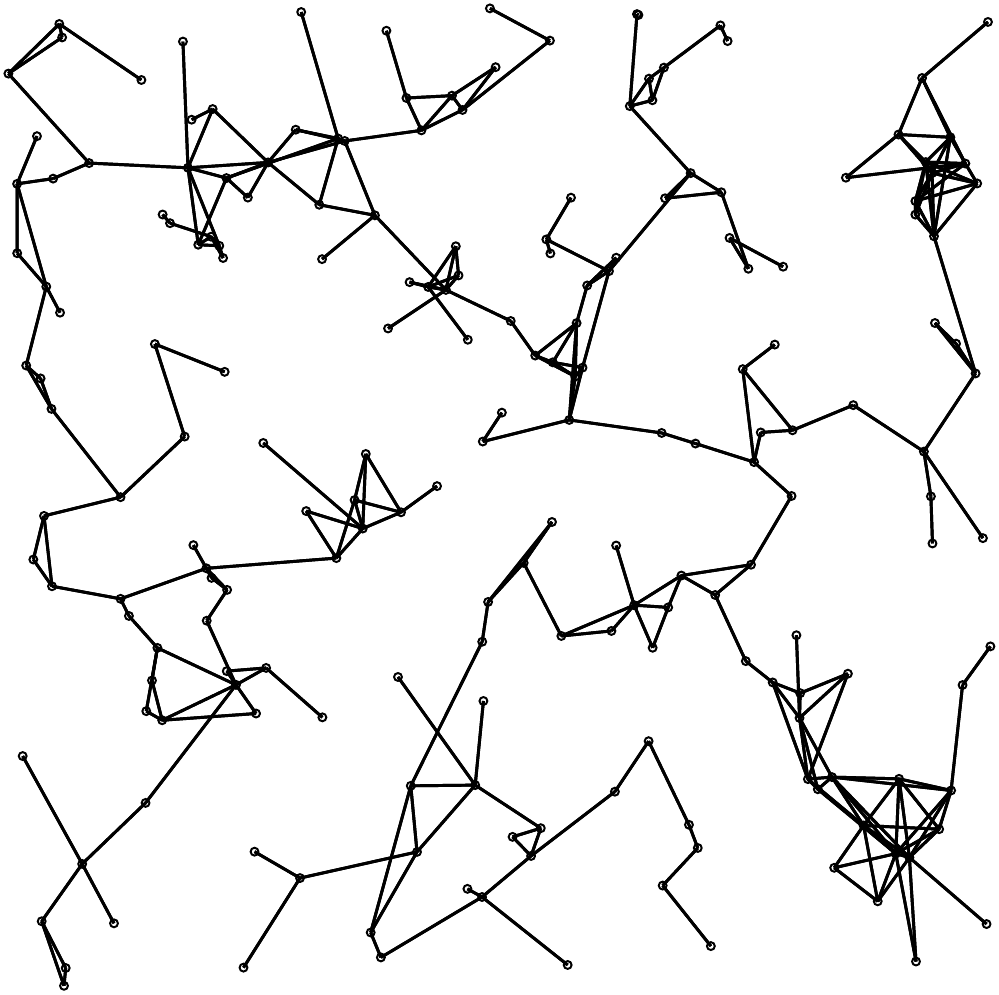}
        }\\

    \caption{Graphs generated by the topology control algorithms when
      the path loss exponents vary around a mean of 3.1 and a standard deviation of
      0.16 (indoor propagation \cite{Rap2002}). For the
      OPT-CBTC($5\pi/6$) algorithm which assumes a 
      uniform path loss exponent in the entire region, the plot shows
      the graph generated when the path loss exponent is 3.1 for all
      node pairs. Figures (a)--(g) depict the graphs generated by
      the respective topology control algorithms while Figures
      (h)--(j) depict the cover graphs. Each network consists of 200
      randomly located nodes in a unit square area.}
\end{center}
\end{figure*}

In this set of experiments, we study the
effectiveness of the algorithms across a five-fold increase in the
number of nodes from 100 to 500. In these experiments, as described
earlier, we use a truncated Gaussian distribution with a mean of 3.1
and a standard deviation of 0.16 based on empirical measurements of
indoor propagation reported in \cite{Rap2002}. Figure
\ref{fig:Fig4_avgOmniPower} shows that the STC algorithm
performs better than other algorithms independent of the number of
nodes as regards the first objective of the topology control algorithm
specified in Section~\ref{problem_statement}. Figure
\ref{fig:Fig4_avgOptPathOmniPower} plots the ratio
$C_{T/H}(u \rightarrow v, {\mathrm Energy})$ averaged over all pairs
of nodes $u$ and $v$ for minimum-energy paths.  

In fact, all of the topology control algorithms in our study
scale similarly since they are all localized algorithms where no
control information propagates beyond more than two hops. The plots
for interference and average node degrees are similarly flat against
the number of nodes for each of the topology control algorithms. All
of the plots indicate that the STC achieves lower energy costs and
interference than other algorithms independent of the number of nodes
in the network. 

Finally, Figures \ref{fig:maxPower}--\ref{fig:MST_Omni} present a
pictorial representation of the graphs generated by the topology
control algorithms. For these figures, as described earlier, we use a
truncated Gaussian distribution with a mean of 3.1 and a standard
deviation of 0.16 based on empirical measurements of indoor
propagation reported in \cite{Rap2002}, except for the case of
OPT-CBTC($5\pi/6$) for which we used the uniform value of 3.1 between
all node pairs (because CBTC algorithms assume that the path loss 
exponents are the same across the entire region of the
network). Figures \ref{fig:maxPower}--\ref{fig:MST} are not the cover
graphs since it is harder to observe the distinct edges and the sparseness
achieved by the algorithms in the more dense cover graphs. Figures
\ref{fig:DLSS_Omni}--\ref{fig:MST_Omni} depict the cover graphs
generated by the DLSS, STC and MST, the ones that generate the
most sparse cover graphs.

\subsection{Experiment 4}

\begin{figure*}[!t]
\begin{center}
    \subfigure[{Average $P_{T/H}(u)$ over all $u$.}]{
        \label{fig:Fig6_avgOmniPower}
        \includegraphics[width=3in]{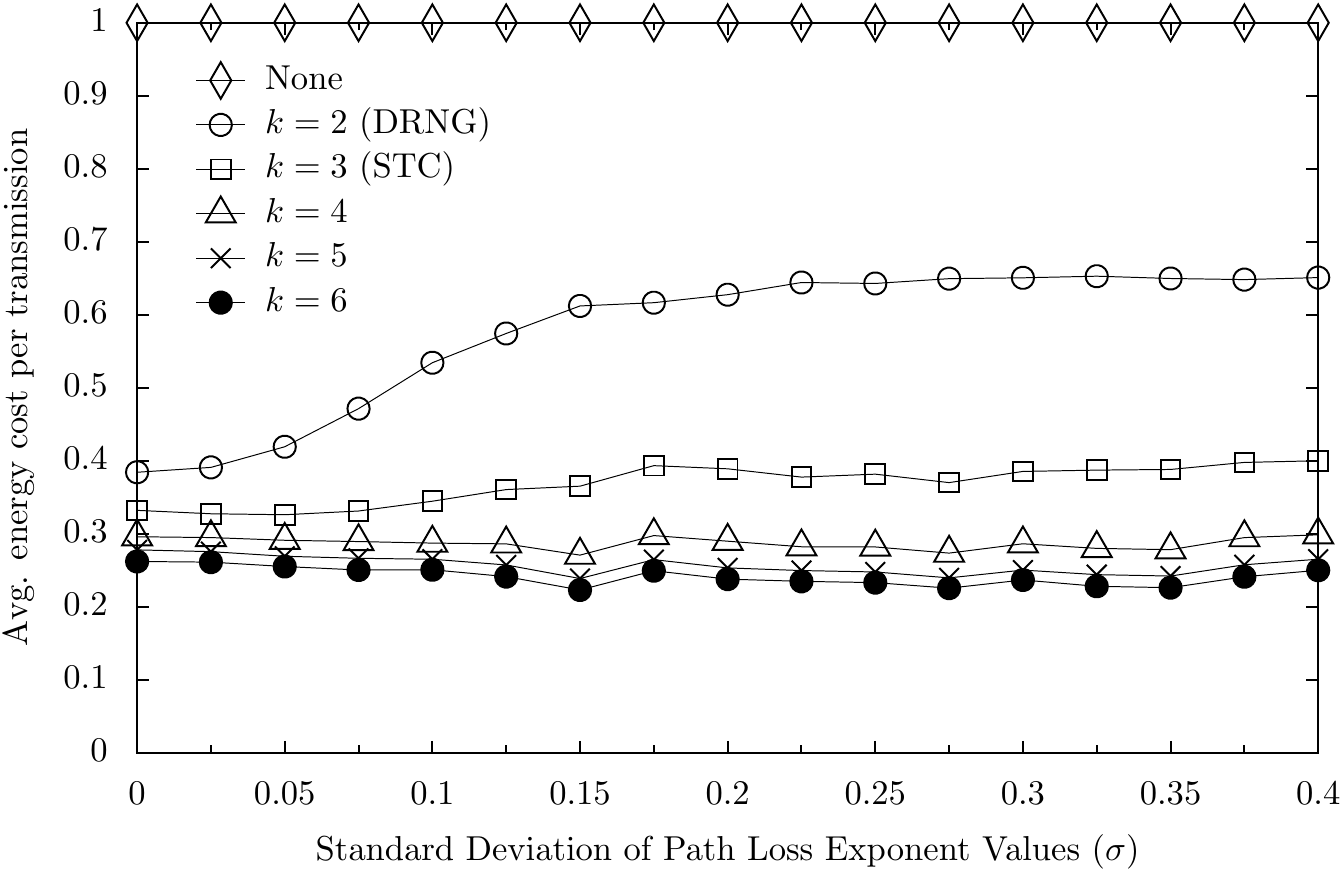}
        }
    \hskip0.2in 
    \subfigure[{$C_{T/H}(u \rightarrow v, {\mathrm Energy})$ averaged over all
      pairs of nodes $u$ and $v$.}]{
        \label{fig:Fig6_avgOptPathOmniPower}
        \includegraphics[width=3in]{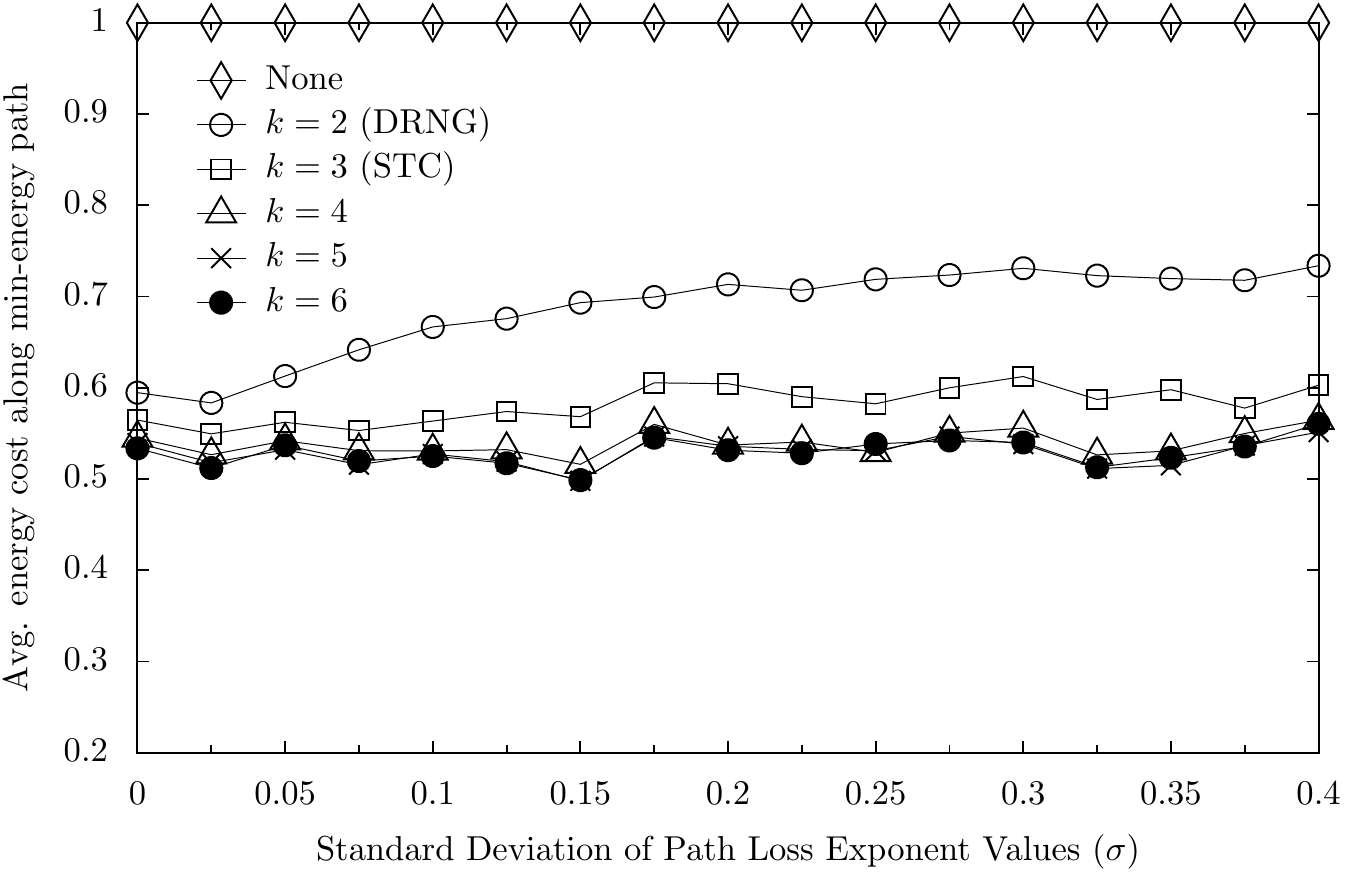}
        }
    \caption{STC and DRNG are specific instances of a class of
      protocols parametrized by a positive integer $k$, in 
      which an edge $(u, v)$ is dropped iff there exists a path of $k$
      hops or less from $u$ to $v$ and from $v$ to $u$, and iff the
      transmission across each of these hops involves an energy cost
      smaller than that for a direct transmission between $u$ and
      $v$. These graphs make comparisons between the energy
      performance achieved for different values of $k$.} 
\end{center}
\end{figure*}

STC and DRNG are specific instances of a class of
protocols parametrized by a positive integer $k$, in which an edge
$(u, v)$ is dropped if there exists a path of $k$ hops or less from
$u$ to $v$ and from $v$ to $u$, and if the transmission across each of
these hops involves an energy cost smaller than that for a direct
transmission between $u$ and $v$. A legitimate question one may wish
to address is if larger values of $k$ will yield better performance
than $k=2$ (DRNG or XTC) and $k=3$ (STC).

Figure \ref{fig:Fig6_avgOmniPower} presents the ratio $P_{T/H}(u)$
averaged over all nodes for values of $k$ from 2 to 6. Figure
\ref{fig:Fig6_avgOptPathOmniPower} plots the ratio $C_{T/H}(u
\rightarrow v, {\mathrm Energy})$ averaged over all pairs of nodes $u$
and $v$ for the same values of $k$. As expected, the average power at
which nodes have to make their transmissions reduces as $k$
increases (although the reductions after $k=3$ are not as significant
as that from $k=2$ to $k=3$). The average energy cost across an
energy-optimal path, however, increases with $k=6$ after reaching its
best values at $k=5$. While the two figures indicate that the best
performance may be achieved at $k=5$ in terms of energy, it would come
at the expense of a significant increase in computational and
communication costs (which themselves may negate the energy savings
with the reduced number of edges). On the other hand, as shown in
Section~\ref{sec:Complexity}, moving from $k=2$ to $k=3$ does not
incur an increase in the communication or computational complexity
while achieving a significant increase in energy savings.

\section{Concluding Remarks}
\label{sec:Conclusion}

Real wireless environments are characterized by a variety of
irregularities and by the phenomenon of multipath propagation. Without 
topology control algorithms in these environments with large
variations in loss characteristics, the nodes in a network may be
forced to use very high power levels in their transmissions to ensure
communication and network connectivity. Most topology control
algorithms do not accommodate for the unique requirements of real
wireless environments and often assume the ability of a node to deduce
spatial information about its neighbors. The Step Topology Control
(STC) algorithm presented in this paper makes no location-based
assumptions and achieves a reduction in the overall energy consumption
without the use of GPS devices or estimations of distance and
direction.  

We have presented simulation results studying the energy consumption
properties of the STC algorithm in comparison to other algorithms that
similarly do not employ location-based information. While the STC algorithm
certainly performs better than other algorithms when the loss
characteristics are uniform in the region of the network, it performs
significantly better when there exists a variation in these loss
characteristics. This makes the STC algorithm especially desirable in
the presence of multipath propagation or when the path loss exponents
are non-uniform in the region of interest. A theoretical analysis of
the energy savings achieved by the STC algorithm in the presence of
varying path loss exponents remains an open problem. 

Most interestingly, the performance advantages of the STC algorithm
come without an increase in the order of communication or
computational complexity. In fact, DLSS, the topology control
algorithm that comes closest to the STC algorithm in performance, has
a higher computational complexity than the STC algorithm. 


We show that the STC algorithm is related to the OPT-CBTC($5\pi/6$)
algorithm and retains some of the angular properties of the CBTC
algorithms. When the path loss characteristics are known to a node
within its neighborhood, these properties may be employed in an
optimization of the discovery process to determine when the local
neighborhood is fully discovered for topology control purposes.

\section*{Appendix}

R{\small ELATIONSHIP TO} C{\small ONE}-B{\small ASED} T{\small
  OPOLOGY} C{\small ONTROL}\\
\label{app:CBTC-STC-relationship}

In the cone-based topology control (CBTC) algorithm, a node $u$ determines the minimum
power $p_{u,\alpha}$ at which it can make an omni-directional broadcast and
successfully reach at least one neighbor node in each cone/sector of angle
$\alpha$. An edge $(u,v)$ is removed from the network topology in an
execution of CBTC($\alpha$) if $u$ cannot reach $v$ with power
$p_{u,\alpha}$ {\em and} $v$ cannot reach $u$ at power
$p_{v,\alpha}$. It is proved in \cite{LiHal2005} that when $\alpha
\leq 5\pi/6$, the network connectivity is preserved. Additional
optimizations to further reduce the energy consumption at each node and
remove more edges are possible and these include:
\begin{itemize}{\leftmargin=5em}
\item {\em Shrink-Back Operation}: If after the execution of CBTC($\alpha$)
at a node $u$ not all cones of angle $\alpha$ contain a neighbor
node (i.e., there is an $\alpha$-gap in the cone coverage as in
the case of nodes at the boundary of the network), node $u$ may
unnecessarily transmit at its maximum power. With this 
optimization, a node transmits at the power at which 
further increasing its transmission power does not increase the cone
coverage.
\item {\em Pairwise Edge Removal}: If there is an edge from $u$ to $v_1$ and
from $u$ to $v_2$, then this operation removes the longer edge if
$\angle v_1uv_2 < \pi/3$, even if there is no edge between $v_1$ and
$v_2$. It is proved in \cite{LiHal2005} that this preserves the connectivity.
\end{itemize}

OPT-CBTC($5\pi/6$) is the cone-based topology control algorithm that 
uses CBTC($5\pi/6$) along with the applicable optimizations of the
shrink-back operation and pairwise edge removal. The CBTC algorithms assume
uniform loss characteristics across the entire region of the network
and also assume that the maximum power $P_u$ of each node $u$ is the
same. We will make the same assumptions in the following to prove the
relationship between STC and OPT-CBTC($5\pi/6$).

Let $G_{\mathrm{CBTC}(5\pi/6)} = ( V, E_{\mathrm{CBTC}(5\pi/6)})$ denote the graph obtained by
CBTC($5\pi/6$). Similarly, let $G_{\mathrm{OPT-CBTC}(5\pi/6)} = ( V, E_{\mathrm{OPT-CBTC}(5\pi/6)})$ denote the graph obtained by OPT-CBTC($5\pi/6$). Let $G=(V,E)$ denote the graph obtained by 
the STC algorithm. Let $d(u,v)$ denote the distance between two nodes
$u$ and $v$. We first restate
the lemma from \cite{LiHal2005} that helps in drawing the relationship
we seek.

\begin{lemma}
\label{lemma:cbtc} 
Any edge $(u, v) \in E_{\mathrm max}$
either belongs to $E_{\mathrm{CBTC}(5\pi/6)}$ or there exist $u^\prime$, $v^\prime
\in V$ such that (a) $d(u^\prime, v^\prime) < d(u,v)$, (b) either
$u^\prime = u$ or $(u,u^\prime) \in E_{\mathrm{CBTC}(5\pi/6)}$, and (c) either
$v^\prime = v$ or $(v,v^\prime) \in E_{\mathrm{CBTC}(5\pi/6)}$. 
\end{lemma} 

The above lemma is proved in \cite{LiHal2005}. We now proceed to prove the
relationship between STC and OPT-CBTC($5\pi/6$) by first proving
the relationship between STC and CBTC($5\pi/6$). 

\begin{lemma}
\label{lemma:STC-CBTC5PIby6}
If an edge $(u,v) \notin E_{\mathrm{CBTC}(5\pi/6)}$, then $(u,v) \notin E$.
\end{lemma}

\begin{proof}
Assume that edge $(u,v) \notin E_{\mathrm{CBTC}(5\pi/6)}$. When an edge $(u,n)
\in E_{\mathrm{CBTC}(5\pi/6)}$ and $(u,v) \notin E_{\mathrm{CBTC}(5\pi/6)}$, we know that
$d(u,v) > d(u,n)$ since $n$ is discovered by $u$ at a
certain power level but $v$ is not discovered by $u$. Therefore, using Lemma
\ref{lemma:cbtc}, we know that there exist nodes $u^\prime$ and
$v^\prime$ such that $d(u^\prime, v^\prime) < d(u,v)$  and in
addition, one of the following three conditions is satisfied: 

{\em Case 1:} $u^\prime = u$, $v^\prime \neq v$, $d(v,v^\prime) <
d(u,v)$. In this case, we have a two-hop path between $u$ and $v$
through $v^\prime$ such that both hops are of distance less than
$d(u,v)$. 

{\em Case 2:} $u^\prime \neq u$, $v^\prime = v$, $d(u,u^\prime) <
d(u,v)$. In this case also, we have a two-hop path between $u$ and $v$
through $u^\prime$ such that both hops are of distance less than
$d(u,v)$.  

{\em Case 3:} $u^\prime \neq u$, $v^\prime \neq v$, $d(u,u^\prime) <
d(u,v)$, and $d(v,v^\prime) < d(u,v)$. We now have a three-hop path between $u$ and $v$
through $u^\prime$ and $v^\prime$ such that each of the three hops is of distance less than 
$d(u,v)$. 

Since there exist nodes such that either a two-hop or a three-hop path
exists between $u$ and $v$ with each hop corresponding to a distance
smaller than $d(u,v)$, $(u,v) \notin E$.

\end{proof}

\begin{theorem}
\label{theo:STC-OPT-CBTC5PIby6}
If $(u,v) \notin E_{\mathrm{OPT-CBTC}(5\pi/6)}$, then $(u,v) \notin E$.
\end{theorem}

\begin{proof}
Since we know from Lemma \ref{lemma:STC-CBTC5PIby6} that STC
removes all the edges that are removed by CBTC($5\pi/6$), we only have
to prove that STC also removes the edges removed by the optimizations
of the shrink-back operation and pairwise edge removal.

When an edge $(u,v)$ is removed as part of the Shrink-Back operation,
the cone coverage around $u$ does not change. Therefore, there exist
nodes $u^\prime$ and $v^\prime$ such that $d(u^\prime, v^\prime) < d(u,v)$  and in
addition, one of the three cases listed in the proof of Lemma
\ref{lemma:STC-CBTC5PIby6} applies. Exactly as in the proof of Lemma \ref{lemma:STC-CBTC5PIby6},
this implies that $(u,v) \notin E$. 

When the Pairwise Edge Removal operation removes an edge $(u,v)$, it
implies there is another neighbor node $(u, n)$ such that $d(u,n)
< d(u,v)$ and $\angle nuv < \pi/3$. Since $\angle nuv < \pi/3$, edge
$(n,v)$ is not the longest edge in the triangle $nuv$. Since $d(u,n)$
is also smaller than $d(u,v)$, we have a two-hop path between $u$ and
$v$ through $n$ where the distance across each hop is less than
$d(u,v)$. Therefore, edge $(u,v)$ is removed by STC as well.
\end{proof}

\section*{Acknowledgment}

This work was supported in part by NSF Awards CNS-0322797 and CNS-0626548. 

\bibliographystyle{IEEEtran}
\bibliography{../../bib/all}

\begin{thebibliography}{10}
\providecommand{\url}[1]{#1}
\csname url@samestyle\endcsname
\providecommand{\newblock}{\relax}
\providecommand{\bibinfo}[2]{#2}
\providecommand{\BIBentrySTDinterwordspacing}{\spaceskip=0pt\relax}
\providecommand{\BIBentryALTinterwordstretchfactor}{4}
\providecommand{\BIBentryALTinterwordspacing}{\spaceskip=\fontdimen2\font plus
\BIBentryALTinterwordstretchfactor\fontdimen3\font minus
  \fontdimen4\font\relax}
\providecommand{\BIBforeignlanguage}[2]{{%
\expandafter\ifx\csname l@#1\endcsname\relax
\typeout{** WARNING: IEEEtran.bst: No hyphenation pattern has been}%
\typeout{** loaded for the language `#1'. Using the pattern for}%
\typeout{** the default language instead.}%
\else
\language=\csname l@#1\endcsname
\fi
#2}}
\providecommand{\BIBdecl}{\relax}
\BIBdecl

\bibitem{San2006}
P.~Santi, \emph{Topology Control in Wireless Ad Hoc and Sensor Networks}.\hskip
  1em plus 0.5em minus 0.4em\relax John Wiley and Sons, 2006.

\bibitem{LiHou2005-1195}
N.~Li, J.~C. Hou, and L.~Sha, ``Design and analysis of an {MST}-based topology
  control algorithm,'' \emph{IEEE Transactions on Wireless Communications},
  vol.~4, no.~3, pp. 1195--1206, May 2005.

\bibitem{RodMen1999}
V.~Rodoplu and T.~H. Meng, ``Minimum energy mobile wireless networks,''
  \emph{IEEE Journal on Selected Areas in Communications}, vol.~17, no.~8, pp.
  1333--1344, August 1999.

\bibitem{LiHal2001}
L.~Li and J.~Y. Halpern, ``Minimum energy mobile wireless networks revisited,''
  in \emph{Proceedings of the IEEE International Conference on Communications
  (ICC)}.\hskip 1em plus 0.5em minus 0.4em\relax IEEE, 2001, pp. 278--283.

\bibitem{JiaRaj2003}
L.~Jia, R.~Rajaraman, and C.~Scheideler, ``On local algorithms for topology
  control and routing in ad hoc networks,'' in \emph{Proceedings of the ACM
  symposium on Parallel Algorithms and Architectures (SPAA)}.\hskip 1em plus
  0.5em minus 0.4em\relax ACM Press, 2003, pp. 220--229.

\bibitem{LiSon2005}
X.-Y. Li, W.-Z. Song, and Y.~Wang, ``Efficient topology control for ad-hoc
  wireless networks with non-uniform transmission ranges,'' \emph{Wireless
  Networks}, vol.~11, no.~3, pp. 255--264, 2005.

\bibitem{LiHal2005}
L.~E. Li, J.~Y. Halpern, P.~Bahl, Y.-M. Wang, and R.~Wattenhofer, ``A
  cone-based distributed topology control algorithm for wireless multi-hop
  networks,'' \emph{IEEE/ACM Transactions on Networking}, vol.~13, no.~1, pp.
  147--159, February 2005.

\bibitem{PucHae2006}
D.~Puccinelli and M.~Haenggi, ``Multipath fading in wireless sensor networks:
  measurements and interpretation,'' in \emph{Proceedings of the 2006
  International Conference on Communications and Mobile Computing}, 2006, pp.
  1039--1044.

\bibitem{ZhoHe2006}
G.~Zhou, T.~He, S.~Krishnamurthy, and J.~A. Stankovic, ``Models and solutions
  for radio irregularity in wireless sensor networks,'' \emph{ACM Transactions
  on Sensor Networks}, vol.~2, no.~2, pp. 221--262, 2006.

\bibitem{LiHou2005-1313}
N.~Li and J.~C. Hou, ``Localized topology control algorithms for heterogeneous
  wireless networks,'' \emph{IEEE/ACM Transactions on Networking}, vol.~13,
  no.~6, pp. 1313--1324, December 2005.

\bibitem{IEEE802.11std}
\BIBentryALTinterwordspacing
{LAN MAN Standards Committee of the IEEE Computer Society}, ``Wireless {LAN}
  medium access control ({MAC}) and physical layer ({PHY}) specifications,''
  2003. [Online]. Available: \url{http://standards.ieee.org}
\BIBentrySTDinterwordspacing

\bibitem{BloLeo2003}
D.~M. Blough, M.~Leoncini, G.~Resta, and P.~Santi, ``The $k$-{N}eigh protocol
  for symmetric topology control in ad hoc networks,'' in \emph{Proceedings of
  the International Symposium on Mobile Ad Hoc Networking and Computing (ACM
  MobiHoc)}.\hskip 1em plus 0.5em minus 0.4em\relax ACM Press, 2003, pp.
  141--152.

\bibitem{WatZol2004}
R.~Wattenhofer and A.~Zollinger, ``{XTC}: A practical topology control
  algorithm for ad hoc networks,'' in \emph{Proceedings of the International
  Parallel and Distributed Processing Symposium (IPDPS)}.\hskip 1em plus 0.5em
  minus 0.4em\relax IEEE, 2004, pp. 216--223.

\bibitem{DamPan2006}
M.~Damian, S.~Pandit, and S.~Pemmaraju, ``Local approximation schemes for
  topology control,'' in \emph{Proceedings of the Annual ACM Symposium on
  Principles of Distributed Computing (PODC)}.\hskip 1em plus 0.5em minus
  0.4em\relax ACM Press, 2006, pp. 208--217.

\bibitem{KuhZol2003}
F.~Kuhn and A.~Zollinger, ``Ad-hoc networks beyond unit disk graphs,'' in
  \emph{Proceedings of the Joint Workshop on Foundations of Mobile Computing
  (DIALM-POMC)}.\hskip 1em plus 0.5em minus 0.4em\relax ACM Press, 2003, pp.
  69--78.

\bibitem{BloLeo2005}
D.~M. Blough, M.~Leoncini, G.~Resta, and P.~Santi, ``Topology control with
  better radio models: {I}mplications for energy and multi-hop interference,''
  in \emph{Proceedings of the 8th ACM International Symposium on Modeling,
  Analysis and Simulation of Wireless and Mobile Systems (MSWiM)}.\hskip 1em
  plus 0.5em minus 0.4em\relax ACM Press, 2005, pp. 260--268.

\bibitem{Kru1956}
J.~B. Kruskal, ``On the shortest spanning subtree of a graph and the traveling
  salesman problem,'' \emph{Proceedings of the American Mathematical Society},
  vol.~7, no.~1, pp. 48--50, February 1956.

\bibitem{SonWan2004}
W.-Z. Song, Y.~Wang, X.-Y. Li, and O.~Frieder, ``Localized algorithms for
  energy efficient topology in wireless ad hoc networks,'' in \emph{MobiHoc
  '04: Proceedings of the 5th ACM international symposium on Mobile ad hoc
  networking and computing}.\hskip 1em plus 0.5em minus 0.4em\relax New York,
  NY, USA: ACM, 2004, pp. 98--108.

\bibitem{Rap2002}
T.~S. Rappaport, \emph{Wireless Communications: Principles and Practice},
  2nd~ed.\hskip 1em plus 0.5em minus 0.4em\relax Prentice Hall, 2002.

\bibitem{JohCar2005}
T.~Johansson and L.~Carr-Moty\v{c}kov\'{a}, ``Reducing interference in ad hoc
  networks through topology control,'' in \emph{Proceedings of the Joint
  Workshop on Foundations of Mobile Computing (DIALM-POMC)}.\hskip 1em plus
  0.5em minus 0.4em\relax ACM Press, 2005, pp. 17--23.

\bibitem{BurRic2004}
M.~Burkhart, P.~von Rickenbach, R.~Wattenhofer, and A.~Zollinger, ``Does
  topology control reduce interference?'' in \emph{Proceedings of the ACM
  International Symposium on Mobile Ad Hoc Networking and Computing
  (MobiHoc)}.\hskip 1em plus 0.5em minus 0.4em\relax ACM Press, 2004, pp.
  9--19.

\bibitem{HayMoh2005}
S.~Haykin and M.~Moher, \emph{Modern Wireless Communications}.\hskip 1em plus
  0.5em minus 0.4em\relax Prentice Hall, 2005.

\bibitem{ErcGre1999}
V.~Erceg, L.~J. Greenstein, S.~Y. Tjandra, S.~R. Parkoff, A.~Gupta, B.~Kulic,
  A.~A. Julius, and R.~Bianchi, ``An empirically based path loss model for
  wireless channels in suburban environments,'' \emph{IEEE Journal on Selected
  Areas in Communications}, vol.~17, no.~7, pp. 1205--1211, 1999.

\bibitem{DabHai2006}
J.~A. Dabin, A.~M. Haimovich, and H.~Grebel, ``A statistical ultra-wideband
  indoor channel model and the effects of antenna directivity on path loss and
  multipath propagation,'' \emph{IEEE Journal on Selected Areas in
  Communications}, vol.~24, no.~4, pp. 752--758, 2006.

\bibitem{LieRei2007}
L.~C. Liechty, E.~Reifsnider, and G.~Durgin, ``Developing the best 2.4 {GHz}
  propagation model from active network measurements,'' in \emph{Proceedings of
  the Vehicular Technology Conference}.\hskip 1em plus 0.5em minus 0.4em\relax
  IEEE, 2007, pp. 894--896.

\bibitem{GenVai2007}
S.~Geng and P.~Vainikainen, ``Experimental investigation of the properties of
  multiband {UWB} propagation channels,'' in \emph{Proceedings of the Annual
  IEEE Symposium on Personal, Indoor and Mobile Radio Communications}.\hskip
  1em plus 0.5em minus 0.4em\relax IEEE, 2007, pp. 1--5.

\bibitem{IraBel2006}
Z.~Irahhauten, G.~Bellusci, G.~J.~M. Janssen, H.~Nikookar, and C.~Tiberius,
  ``Investigation of {UWB} ranging in dense indoor multipath environments,'' in
  \emph{Procedings of the IEEE Singapore International Conference on
  Communication systems}.\hskip 1em plus 0.5em minus 0.4em\relax IEEE, 2006,
  pp. 1--5.

\end{thebibliography}

\end{document}